# Collective behavior of active topological solitons, knotted streamlines, and transport of cargo in liquid crystals

Hayley R. O. Sohn,[1] Paul J. Ackerman,[2] Timothy J. Boyle,[2] Ghadah H. Sheetah,[1] Bengt Fornberg,[3] and Ivan I. Smalyukh[1,2,4*]

[1]Soft Materials Research Center and Materials Science and Engineering Program, University of Colorado, Boulder, CO 80309, USA
[2]Department of Physics and Department of Electrical, Computer and Energy Engineering, University of Colorado, Boulder, CO 80309, USA
[3]Department of Applied Mathematics, University of Colorado, Boulder, CO 80309, USA
[4]Renewable and Sustainable Energy Institute, National Renewable Energy Laboratory and University of Colorado, Boulder, CO 80309, USA.

*ivan.smalyukh@colorado.edu

*Active colloids and liquid crystals are capable of locally converting the macroscopically-supplied energy into directional motion and promise a host of new applications, ranging from drug delivery to cargo transport at the mesoscale. Here we uncover how topological solitons in liquid crystals can locally transform electric energy to translational motion and allow for the transport of cargo along directions dependent on frequency of the applied electric field. By combining polarized optical video microscopy and numerical modeling that reproduces both the equilibrium structures of solitons and their temporal evolution in applied fields, we uncover the physical underpinnings behind this reconfigurable motion and study how it depends on the structure and topology of solitons. We show that, unexpectedly, the directional motion of solitons with and without the cargo arises mainly from the asymmetry in rotational dynamics of molecular ordering in liquid crystal rather than from the asymmetry of fluid flows, as in conventional active soft matter systems.*

# I. INTRODUCTION.

For nearly two centuries, in fields of physics ranging from fluid dynamics to optics, rich dynamic behavior of self-reinforcing solitary wave packets has attracted a great deal of interest among physicists and mathematicians alike [1,2]. These solitons maintain their spatially-localized shape while propagating and typically emerge from a delicate balance of nonlinear and dispersive effects in the physical host medium [1]. Solitons of a very different type, often called "topological solitons", are topologically-nontrivial, spatially-localized nonsingular field configurations that are rarely associated with out-of-equilibrium dynamics, but rather are studied as static field configurations embedded in a uniform background [3]. Their topologically-nontrivial configurations can be classified using homotopy theory [3], though their stability in real physical systems usually also requires nonlinearities [3-17]. For example, in particle physics, topological solitons called "skyrmions" are unstable within the linear models [3,5,6], but can be stabilized by adding nonlinear terms [3,4,7]. In non-centrosymmetric ferromagnets and chiral liquid crystals (LCs), various two-dimensional (2D) and three-dimensional (3D) condensed matter counterparts of these topological solitons are stabilized by the medium's tendency to form twisted field configurations [9-17], which is associated with additional nonlinear terms added to the harmoniclike free-energy potential (e.g., the Dzyaloshinskii-Moriya term in the case of chiral ferromagnets). However, unexpected recent observations show that ferromagnetic skyrmions can move over large distances within solid thin films while maintaining their topology of localized nonsingular spin textures [18,19]. These findings may one day enable race track memory devices for improved information storage and drive a great deal of interest for potential applications in spintronics, so that even a new term "skyrmionics" has been coined [19]. In LCs, topological solitons realized in localized configurations of the molecular alignment field (describing the spatial orientation pattern of rodlike molecules), called the "director field" **n(r)**, can be moved by applying external fields, though this motion of the soliton now emerges in a fluid medium rather than in a solid film [20], which may potentially cause a rich interplay between the motion of the localized field configuration and LC fluid flows.

In this work, we realize reconfigurable active motion of various topologically-nontrivial skyrmionic and knotted field configurations in chiral nematic LCs. Like singular defects in active matter [21-24], topological solitons exhibit directional motion both as individual objects and collectively, often spontaneously selecting and synchronizing their motion directions as this out-of-equilibrium process progresses. However, unlike in the case of singular active matter defects [21-24], this motion is not accompanied with annihilation and generation of defects, can persist for months, and its direction can be controllably reversed. By using a combination of optical microscopy and 3D modeling of both the equilibrium free-energy-minimizing director structures and their temporal evolution, we uncover the physical mechanisms behind the soliton motion. We demonstrate that this motion emerges from spatially-asymmetric changes of director structures that evolve very differently and non-reciprocally upon the application and removal of an electric field, so that the periodic modulation of an applied field yields net translational motion of solitons. This asymmetry of back-forth cycles of the topology-constrained director field evolution resembles squirming of biological cells and biomimetic robots [25,26], though it is also very different and unique in that it mainly involves the rotational dynamics of the director field configuration with little or no coupling to the LC fluid flows. The modulated high-frequency electric field in our experiments is applied to the entire sample with volume $10^8$-times larger than that occupied by a typical single solitonic structure. The facile response of the LC results in the strong coupling between the electric field **E** and director **n**, causing a periodic local conversion of electric energy into the elastic energy stored within the distorted soliton and then into the soliton's translational squirming motion. This behavior, exhibited by several different species of topological LC solitons such as skyrmions, torons, and cholesteric fingers, reveals a novel type of solitonic active soft matter that may potentially find technological uses. Our numerical modeling adopts an approach based on using meshless 3D node sets [27,28], allowing us to reproduce fine details of the equilibrium director structures of the solitons (such as the voltage-dependent 3D knotted streamlines of the LC molecular alignment field), as well as their temporal evolution. Finally, we show how the anisotropic interactions of the director field with surfaces of colloidal particles can be used for entrapping them by solitons, as well as for transporting these micrometer- and nanometer-sized cargo particles.

## II. MATERIALS AND METHODS.

**II.A. Material and sample preparation.** Chiral nematic LC mixtures with both positive and negative dielectric anisotropy $\Delta\varepsilon$ are prepared by mixing a room-temperature nematic host with a chiral additive. According to the relation $p_0 = 1/(h_{HTP} \cdot c)$ [9,20], an LC mixture with the ground-state helicoidal pitch $p_0$ is prepared by controlling the concentration of the chiral additive $c$ for the known helical twisting power $h_{HTP}$ of the chiral additive in the nematic host (Table 1).

Glass substrates with transparent indium tin oxide (ITO) electrodes were treated with polyimide SE-1211 (purchased from Nissan) to impose strong homogeneous vertical surface boundary conditions. The SE-1211 was applied to the ITO electrodes via spin-coating at 2700 rpm for 30 s then baked for 5 min at 90°C and 1 h at 190°C to induce cross-linking of the alignment layer. Glass fiber segments dispersed in ultraviolent-curable glue were used as spacers to set the inter-substrate separation gap, $d = 10 - 60$ μm. Small drops of the glue with spacers were sandwiched between the substrates with ITO electrodes and alignment layers facing inward and cured for 60 s using ultraviolent exposure (OmniCure UV lamp, Series 2000). Leads were soldered to the ITO electrodes to provide electrical connection for application of an electric potential across the LC cell. The chiral nematic LC materials were infiltrated into the confinement cells by means of capillary forces and the edges of the cells were sealed with 5-min fast-setting epoxy.

To obtain the various topological solitons in materials with negative $\Delta\varepsilon$, we used a mixture of the nematic host MLC-6609 or ZLI-2806 and the chiral additive ZLI-811 (all purchased from Merck) with equilibrium helicoidal pitch tuned to $p_0 \approx 10$ μm. For the study of skyrmions in a positive $\Delta\varepsilon$ material, we used a mixture of the nematic host E7 and the chiral additive CB-15 (both purchased from EM Industries) with equilibrium helicoidal pitch $p_0 \approx 30$ μm. As local and global minima of free energy, the solitons were occurring spontaneously in cells with $d/p_0 \approx 1$, as well as were controllably generated using laser tweezers, as detailed below.

**II.B. Voltage modulation.** Homemade MATLAB-based software coupled with a data acquisition board (NIDAQ-6363, National Instruments) was used to produce various voltage-driving schemes and waveforms, which were then applied to the LC cells using wires soldered to the ITO electrodes. In order to avoid hydrodynamic instabilities and other types of complex behavior associated with ion motion induced by low-frequency applied fields, a square carrier waveform with a relatively high frequency of $f_c$ = 1 kHz was used throughout. The voltage-driving scheme utilizes amplitude modulation of the carrier waveform by a square wave at a lower frequency, $f_m$. The homemade software allowed us to tune the frequencies, amplitude of voltage applied ($V_{rms}$), duty cycle (percentage of high carrier amplitude), and fill ratio (with 0 corresponding to no applied voltage during the fraction of a period in which the signal carrier amplitude is low).

**II.C. Tracer nanoparticles.** Gold nanorods (GNRs) used in our study have an aspect ratio of about 4, with dimensions of 109 x 28 nm. We synthesized these GNRs following a seed-mediated method with the adjustment of binary surfactants and pH control [29, 30]. To produce the seed in a glass bottle, 5 mL of hexadecyltrimethylammonium Bromide (CTAB, Sigma-Aldrich, 0.2 M) was added to 5 mL Gold(III) Chloride trihydrate (HAuCl4.3H2O, Sigma-Aldrich, 0.5 mM), followed by a quick addition of 0.6 mL of freshly-prepared, ice-cold Sodium Borohydride (NaBH4, Sigma-Aldrich, 10 mM). The seed was stirred vigorously for 2 min then left at room temperature for 30 min to ensure the decomposition of Sodium Borohydride was fully oxidized. The growth solution was prepared by mixing 247 mg of Sodium Oleate (NaOL, TCI America) with 50 mL of CTAB (77 mM) in a clean flask. The solution was kept around 50 °C, stirred until dissolved, and cooled to room temperature. Once cooled, 3.6 mL of Silver Nitrate (AgNO3, Sigma-Aldrich, 4 mM) was added and left undisturbed for 15 min. 50 mL of HAuCl4 (1 mM) was added and stirred at 700 rpm for 90 min, after which the solution became colorless and 420 µL Hydrochloric Acid (HCl 37 wt.% in water, Fisher Scientific, 12.1 M) was added. The mixture was stirred at 500 rpm for 15 min then 250 µL Ascorbic Acid (Sigma-Aldrich, 80 mM) was added during vigorous stirring for 30 s followed by 80 µL of the seed. The solution was kept undisturbed at room temperature for 12 h before it was centrifuged at 7000 rpm for 20 min. The solution was then centrifuged twice at 7000 rpm for 20 min to remove excess additives and CTAB, after which there was minimal CTAB concentration (2 mM) in the GNR solution, as needed to prevent aggregation.

To disperse the GNRs in an organic medium, we performed surface functionalization through a ligand exchange. A thiol-terminated methoxy-poly(ethylene glycol) (mPEG-SH 5kDa, JemKem Technology) [31] was used, as described in detail in our previous study [32]. A 1 mL aqueous suspension of 30 mg PEG was added to a 50 mL diluted dispersion of GNRs with an optical density of 4. This mixture was left for 12 h and then centrifuged for 10 min at 7000 rpm. The exchange process was repeated and the particles were ready to use after washing them with methanol 2-3 times.

To disperse the GNRs in MLC-6609, we first dispersed them in nematic 4-cyano-4'-pentylbiphenyl (5CB, Chengzhi Yonghua Display Materials Co. Ltd). 30 µL of the GNR dispersion was added in a 0.5 mL centrifuge tube and the solvent was left for an hour at 50°C to fully evaporate. Then 15 µL of 5CB was added and the mixture was kept under sonication at 40°C for 5 min, followed by vigorous stirring until well dispersed in the nematic phase. To transfer the GNRs to MLC-6609 and obtain dilute dispersions of tracer particles, we added one part in 50 of 5CB containing GNRs into MLC-6609 to maintain the negative dielectric anisotropy of the MLC-6609.

In addition to nanorods, we utilized semiconductor nanocubes with average dimensions of about 22 nm, which were synthesized by following procedures described elsewhere [33]. When excited with a 980 nm infrared laser [34], these nanoparticles exhibit strong photon up-converting luminescence, which is then used to track nanoparticle positions and to probe the potential fluid flows by means of video microscopy.

**II.D. Generation of twisted solitons with laser tweezers.** Some of the 3D solitonic structures were controllably "drawn" in the LC cells using optical tweezers comprised of a 1064 nm Ytterbium-doped fiber laser (YLR-10-1064, IPG Photonics) and a phase-only spatial light modulator (P512-1064, Boulder Nonlinear Systems). This setup is capable of controllably producing arbitrary, dynamically-evolving 3D patterns of laser light intensity within the sample [10]. We generated 3D structures by means of optically-induced local reorientation, a process in which the LC director couples to the optical-frequency electric field of the laser beam and realigns away from the far-field background **n**$_0$ [9]. We holographically generated patterns of the trapping laser beam's intensity and concisely controlled motion of individual focused laser beams along linear and circular trajectories, including both Gaussian and Laguerre-Gaussian beams. This technique enabled the generation of complex patterns and distortions in the director field that relaxed into global or local elastic free-energy minima, typically fulfilling the chiral material's preference to twist and thus yielding the twisted solitons that we study.

**II.E. Polarizing optical imaging and video microscopy.** We used an Olympus IX-81 inverted microscope, equipped with crossed polarizers, to capture transmission-mode polarizing optical microscopy (POM) images and videos, which were recorded with a charge-coupled device camera (Flea, PointGrey). Various objectives were used to obtain the POM images of solitons, including 10x, 20x, and 50x dry objectives with numerical aperture ranging within $NA = 0.3 - 0.9$. To track the tracer nanoparticle motions, dark-field images were obtained using a dark-field condenser and a 100x oil-immersion objective with an adjustable numerical aperture of $NA = 0.6 - 1.3$. The soliton dynamics and nanoparticle motion were analyzed by processing optical image sequences and videos using the open-source software ImageJ's (National Institute of Health) particle tracking capabilities. The extracted data sets with particle positions within each frame were used to calculate the net displacements, velocities, and speed anisotropies of the solitons and tracer nanoparticles.

**II.F. Numerical methods.**

***II.F.1. Director relaxation method.*** The bulk LC free-energy density $f$ can be described by the Frank-Oseen potential supplemented by the electric field coupling term $f_e$:

$$f = \frac{1}{2}\left[K_{11}(\nabla \cdot \mathbf{n})^2 + K_{22}\left[\mathbf{n} \cdot (\nabla \times \mathbf{n}) + q_0\right]^2 + K_{33}\left[\mathbf{n} \times (\nabla \times \mathbf{n})\right]^2\right] + f_e \qquad (1)$$

where $K_{11}$, $K_{22}$, and $K_{33}$ are splay, twist, and bend elastic constants, respectively, $q_0 = 2\pi/p_0$ is the chiral wave number of the ground-state chiral nematic mixture. The vectorial representation assumes that the scalar order parameter is independent of coordinates. The free-energy density contribution due to an applied electric field is expressed in terms of the local displacement field, $\mathbf{D} = \varepsilon_0 \bar{\bar{\varepsilon}}(\mathbf{r})\mathbf{E}$, and electric field $\mathbf{E}$ where $\varepsilon_0$ is the permittivity of free space, $\bar{\bar{\varepsilon}}(\mathbf{r})$ is the dielectric tensor that can be expressed with index notation in terms of components of the director field as $\varepsilon_{ij} = \varepsilon_0\left(\varepsilon_\perp \delta_{ij} + \Delta\varepsilon n_i n_j\right)$. The electric field coupling term of free-energy density then reads:

$$f_e = \frac{1}{2}\mathbf{D} \cdot \mathbf{E} \qquad (2)$$

We first assume fixed charge and allow the director field to relax, which is then followed by updating the electric field while enforcing the condition $\nabla \cdot \mathbf{D} = 0$ within the spatially-varying dielectric LC material. The derivatives are calculated by a second-order centered finite difference (FD) method or a radial basis function-generated FD method (RBF-FD) described below while using a Cartesian coordinate system. In equilibrium, the free energy is minimized, yielding Euler-Lagrange equation for the director components. Therefore, the relaxation method for the director structure simulations is based on an update formula for the director:

$$n_i^{new} = n_i^{old} - \frac{\Delta t}{\gamma_1}[f]_{n_i} \qquad (3)$$

with the functional derivatives given by:

$$[f]_{n_i} = \frac{\partial f}{\partial n_i} - \frac{d}{dx}\left(\frac{\partial f}{\partial n_{i,x}}\right) - \frac{d}{dy}\left(\frac{\partial f}{\partial n_{i,y}}\right) - \frac{d}{dz}\left(\frac{\partial f}{\partial n_{i,z}}\right) \qquad (4)$$

where $\gamma_1$ is the material's rotational viscosity and $\Delta t$ is the numerical time step. The maximum stable time step $\Delta t_{max} = \gamma_1 h_{min}^2 / (2K_{33})$ is estimated using the material parameters of studied LCs (Table 1) and the minimum node spacing, $h_{min}$. The update is applied iteratively to the director field, re-normalizing it each time to assure $|\mathbf{n}| = 1$. To determine convergence for equilibrium calculations, we monitor the evolution of the space-averaged rate of change of the functional derivatives. This value should approach zero when equilibrium is reached and is used as a stopping criterion for the simulation ($1 \times 10^{-10}$ was used as the equilibrium stopping criterion in the present study).

*II.F.2. Radial Basis Function concept.* Radial Basis Functions (RBFs) provide a powerful numerical methodology for solving partial differential equations, which has many advantages as compared to other approaches when applied to large-scale problems. This method has been used in a diverse range of fields, including fluid mechanics, astrophysics and geosciences, mathematical biology, and computational electromagnetics [27, 28, 35-41]. RBF-FD is unique in combining all of the following features: (1) numerical stability, even when using explicit time stepping of purely convective problems on irregular node layouts; (2) accuracy levels comparable to those of pseudo-spectral and global RBF methods while relying on local approximations; (3) easy local (adaptive) refinements, (4) geometrical flexibility and (5) excellent opportunities for large-scale parallel computing.

Before introducing our approach for the finite difference calculations used in the director relaxation method, let us recall the key concepts of RBF. On a scattered node set with node locations $\mathbf{x}_k$ labeled by an index $k=1 \rightarrow n$, a radially symmetric function, such as a Gaussian of the form $\varphi(r) = e^{-(\varepsilon r)^2}$, can be used as a basis for computing an interpolant by centering $\varphi$ at each node $\varphi(\|\mathbf{x}-\mathbf{x}_k\|)$. Here $\|\mathbf{x}-\mathbf{x}_k\|$ represents the standard Euclidean distance and $\varepsilon$ is a shape parameter. The RBF interpolant can be expressed in the form:

$$s(\mathbf{x}) = \sum_{k=1}^{n} \lambda_k \varphi(\|\mathbf{x}-\mathbf{x}_k\|) \tag{5}$$

Solutions to a linear system of equations determine the unique weights $\lambda_k$ for the interpolating function where $f_k$ is the data value at the $k^{\text{th}}$ node.

$$\begin{bmatrix} \varphi(\|\mathbf{x}_1-\mathbf{x}_1\|) & \varphi(\|\mathbf{x}_1-\mathbf{x}_2\|) & \cdots & \varphi(\|\mathbf{x}_1-\mathbf{x}_n\|) \\ \varphi(\|\mathbf{x}_2-\mathbf{x}_1\|) & \varphi(\|\mathbf{x}_2-\mathbf{x}_2\|) & \cdots & \varphi(\|\mathbf{x}_2-\mathbf{x}_n\|) \\ \vdots & \vdots & \ddots & \vdots \\ \varphi(\|\mathbf{x}_n-\mathbf{x}_1\|) & \varphi(\|\mathbf{x}_n-\mathbf{x}_2\|) & \cdots & \varphi(\|\mathbf{x}_n-\mathbf{x}_n\|) \end{bmatrix} \begin{bmatrix} \lambda_1 \\ \lambda_2 \\ \vdots \\ \lambda_n \end{bmatrix} = \begin{bmatrix} f_1 \\ f_2 \\ \vdots \\ f_n \end{bmatrix} \qquad (6)$$

For many choices of $\varphi$, this system is guaranteed to be nonsingular no matter how any number of distinct nodes are distributed in any number of dimensions. This fact removes issues of singularity of the left-hand side matrix (the $A$-matrix) faced by pseudo-spectral methods when applied to two- and higher-dimensional node sets. The RBF concept can be easily illustrated in 2D (Fig. 1a-c). A scattered node set in 2D with random fluctuations in a scalar value is represented by points with sticks of varying lengths extending to an x-y plane. A unique set of weights applied to Gaussians centered at each node forms the interpolating function, which can be used to construct a surface that conforms to the scattered data. We can generalize some of the node distributions that are suitable to use for RBF to 3D to include periodic lattices, meshes, scattered nodes sets, or even hybrids of these distribution schemes (Fig. 1d-g). While lattice distributions (Fig. 1 d-e) have the advantage of enabling periodic boundary conditions, using a tetrahedral mesh (Fig. 1f) or a scattered node set (Fig. 1g) allows for complex geometries and adaptive refinement. This becomes especially important for calculations in 3D volumes due to their inhibitive computational cost. Taking advantage of this power of RBF, we use different meshes for computation and visualization of data. Irregular node sets are also beneficial to our numerical calculations because they do not impose artificial symmetries which introduce undesired numerical artifacts. Furthermore, RBF methods applied to these irregular node sets are easily parallelized and our implementation provides higher-accuracy calculations for no additional computational time. In this work, we use a hybrid square periodic method with tetrahedral meshes which demonstrates the potential for adaptive schemes and the elegant simplicity of RBF methods on complex node distributions. Not only is the RBF method compatible with all of these node sets without modification, there is no need for special treatment of boundaries for stability of our numerical calculations.

***II.F.3. Radial Basis Function Finite Difference (RBF-FD) method in 3D.*** Our numerical modeling of director structures and their dynamics in LC cells utilizes both random and regular node layouts. In the RBF-FD approach, RBFs are used to supplement polynomials when generating weights in localized scattered node finite-difference-like stencils. Radial functions of the form $\varphi(r) = r^3, r^5, r^7$ are then particularly effective [42,43]. As mentioned above, calculations performed on irregular node layouts enabled by RBF-FD methods permit local (adaptive) node refinements as well as convenient ways of efficiently discretizing irregular geometries. Simulation node sets were generated using a *Quality Tetrahedral Mesh Generator* [44] that has adaptive node refinement support and additional square periodic padding to implement periodic boundaries along lateral directions.

For traditional grid-based FD calculations, derivatives can be approximated with a single stencil, corresponding set of weights at all nodes, and only minor modification near the boundaries. With scattered nodes, however, each stencil and corresponding weight becomes different. Computationally-efficient approaches, such as MATLAB's function *knnsearch*, can be implemented to find each node's nearest $n-1$ neighbors such that each node has an associated $n$-node stencil. The weights for each stencil for linear operators (nine first- and second-order derivatives in 3D) are determined by solving linear systems of equations augmented with some additional low-order polynomial terms. For example, in the linear 3D case:

$$\begin{bmatrix} & & & | & 1 & x_1 & y_1 & z_1 \\ & A & & | & \vdots & \vdots & \vdots & \vdots \\ & & & | & 1 & x_n & y_n & z_n \\ - & - & - & + & - & - & - & - \\ 1 & \cdots & 1 & | & & & & \\ x_1 & \cdots & x_n & | & & 0 & & \\ y_1 & \cdots & y_n & | & & & & \\ z_1 & \cdots & z_n & | & & & & \end{bmatrix} \begin{bmatrix} \lambda_1 \\ \vdots \\ \lambda_n \\ - \\ \lambda_{n+1} \\ \lambda_{n+2} \\ \lambda_{n+3} \\ \lambda_{n+4} \end{bmatrix} = \begin{bmatrix} L\varphi(\|\mathbf{x}-\mathbf{x}_1\|)|_{\mathbf{x}=\mathbf{x}_c} \\ \vdots \\ L\varphi(\|\mathbf{x}-\mathbf{x}_n\|)|_{\mathbf{x}=\mathbf{x}_c} \\ - \\ L1|_{\mathbf{x}=\mathbf{x}_c} \\ Lx|_{\mathbf{x}=\mathbf{x}_c} \\ Ly|_{\mathbf{x}=\mathbf{x}_c} \\ Lz|_{\mathbf{x}=\mathbf{x}_c} \end{bmatrix}. \quad (7)$$

Here $A$ represents the $A$-matrix used above in the direct approach for calculating weights for the interpolant, $L$ is a linear operator, and the entries $\lambda_{n+1}$ through $\lambda_{n+4}$ in the solution vector can be ignored. The RBF used in our work is a polyharmonic spline (PHS) of the form $r^{2m-1}$, $m \in \mathbb{N}$. There are additionally $\binom{\rho+D}{\rho}$ polynomial terms up to degree $\rho$ in dimension $D$. The benefits of PHS over other choices for the RBF include numerical stability along boundaries, which eliminates the need for special treatment such as generating ghost nodes or selecting a shape parameter (as in the case of the Gaussian RBF) [41].

## III. RESULTS.

**III.A. Solitons with both skyrmionlike and Hopf and Seifert Fibration features.** A chiral nematic LC with a ground-state pitch $p_0$ confined by substrates treated to enforce strong vertical alignment is in a frustrated geometry. It would prefer to form the ground-state helical structure which is, however, incompatible with the imposed vertical surface boundary conditions. When the separation $d$ of the confining substrate planes is approximately equal to $p_0$, numerous spatially-localized solitonic configurations can be observed, embedded in the frustrated, unwound uniform far field $\mathbf{n}_0$. These long-term metastable or ground-state solitonic field configurations incorporate energetically-favorable twist while meeting the imposed vertical surface boundary conditions and can be controllably generated or removed using laser tweezers.

We start with the topologically-nontrivial skyrmionic field configuration known as an elementary toron [9, 17, 20, 45], which we obtain as a result of numerical free-energy minimization using both the RBF-FD and conventional FD (Fig. 1a-d) methods. In the cell midplane between confining substrates, the elementary toron embeds a π-twist of the director field $\mathbf{n}(\mathbf{r})$ radially from the center in all directions, so that it smoothly meets the uniform $\mathbf{n}_0$ in the soliton's periphery. This skyrmionic configuration is embedded in the bulk LC and terminated by two singular point defects located near the confining substrates (Fig. 2b). To show the nontrivial skyrmion topology of this axially-symmetric structure, we use arrows colored by their polar angle to represent it (Fig. 2c). The far field $\mathbf{n}_0$ (blue) is assumed to correspond to the north pole of the order parameter space of unit vector orientations, the two-sphere $\mathbb{S}^2$ (inset). The vectors in this cross-section of the toron can be mapped to fully cover $\mathbb{S}^2$ once, indicating that the structure corresponds to an elementary skyrmion. This skyrmion, however, is terminated at point defects that match it to the uniform boundary conditions at confining surfaces (Fig. 2b,d), as visualized using isosurfaces corresponding to different values of the z-component of $\mathbf{n}(\mathbf{r})$ (blue, red) and the isosurface depicting the small region of point singularities at which vectors with different orientations meet (gray).

We induce topology-preserving distortions in the toron configuration by applying voltage $U$ to the transparent electrodes, with the ensuing electric field **E** perpendicular to the cell midplane. For a material with a negative dielectric anisotropy, the LC tends to respond by reorienting perpendicular to **E**, competing with the elastic energy to yield a morphed structure that minimizes the total free energy. In the far field, at voltages above a well-defined threshold [20], the field-induced director tilt is accompanied by energy-reducing twist of **n(r)** around a vertical helical axis. This yields the so-called "translationally invariant configuration" (TIC) and a tilted or in-plane orientation of the far-field director in the cell midplane. To embed the skyrmionlike configuration of the toron in the TIC, its structure morphs and the region with the vertical orientation of **n(r)** becomes localized and manifests itself as a nonsingular umbilical region corresponding to the north-pole preimage. The preimages of the skyrmion's north and south poles are separated by π-twist, as before. The skyrmion configuration no longer has axial symmetry, but its topology remains unchanged because mapping **n(r)** to $\mathbb{S}^2$ covers it once, though the far-field director is now oriented along the y-axis in the cell midplane (Fig. 2e). The TIC-embedded elementary toron can be visualized in 3D (Fig. 2f), similar to the case at no applied fields, demonstrating a highly asymmetric structure in which the north-pole preimage wraps partially around the south-pole preimage. The side on which the north-pole preimage comes to rest in equilibrium is determined by the orientation of the midplane tilt, which for cells with initially-perpendicular boundary conditions is selected spontaneously [20, 46].

The detailed analysis of **n(r)** near singular defects shows how these point singularities terminate the skyrmion near confining substrates (Fig. 2g-j). Both top and bottom defects are self-compensating elementary hyperbolic hedgehogs of opposite charge in the vectorized **n(r)**. Within the homotopy theory, the elementary skyrmions and the singular point defects in the vectorized **n(r)** are classified as elements of $\pi_2(\mathbb{S}^2)=\mathbb{Z}$ (or $\pi_2(\mathbb{S}^2/\mathbb{Z}^2)=\mathbb{Z}$ for the nonpolar **n(r)** )[9, 20]. It is therefore natural that the elementary skyrmion tube orthogonal to the cell substrates is terminated by the two $\pi_2(\mathbb{S}^2)=\mathbb{Z}$ point singularities near the confining substrates with strong boundary conditions for **n(r)**. This is consistent with the notion that the spatial translation of a $\pi_2(\mathbb{S}^2)=\mathbb{Z}$ point singularity can leave a trace of the $\pi_2(\mathbb{S}^2)=\mathbb{Z}$ nonsingular topological soliton when this trace is smoothly embedded in the far-field background [47].

As pointed out previously [45], torons exhibit structural features that bring about resemblance of not only skyrmions, but also the mathematical Hopf and Seifert fibrations. The latter can be seen by visualizing the 3D structure of streamlines tangent to **n(r)** and originating from locations in the vertical x-z plane at different distances away from the toron's circular axis (Fig. 2k). These streamlines are found to form various torus knots, like the ones found in toroidal drops of DNA [48,49]. Thus, the regions near the circular axis of the toron resemble fragments of stereographic projection of $\mathbb{S}^3$ to $\mathbb{R}^3$, as in Hopf and Seifert fibrations. Similar to the case of toroidal drops of DNA, this structure of **n(r)** emerges to implement the LC's tendency to twist while forming an axially-symmetric configuration. However, unlike in the case of DNA and other biopolymers, our toron structures emerge in a medium formed by small molecules. As a result, the rate of twist of **n(r)** changes smoothly as one moves away from the toron's circular axis, where the director accommodates the effects of confinement and presence of the singular point defects, so that we observe different torus knots formed by the streamlines (Fig. 2k and Fig. 3).

The 3D director twist is inherently geometrically frustrated due to the fact that it is incompatible with Euclidian 3D space, $\mathbb{R}^3$ [50]. However, one can consider the geometry and topology of fiber bundles to understand how LC can efficiently embed nearly-uniform 3D twist to a torus-interior volume [50]. To describe these fibered spaces, it is convenient to start with a parameterization known as the toroidal coordinates of $\mathbb{S}^3$:

$$x_1 = R\cos(\theta)\sin(\phi)$$
$$x_2 = R\sin(\theta)\sin(\phi)$$
$$x_3 = R\cos(\omega)\cos(\phi)$$
$$x_4 = R\sin(\omega)\cos(\phi)$$

where $x_1$, $x_2$, $x_3$, and $x_4$ are the hyperspherical coordinates in $\mathbb{R}^4$, $R$ is the radius of $\mathbb{S}^3$, $\theta \in [0, 2\pi)$, $\phi \in [0, \pi/2]$, and $\omega \in [0, 2\pi)$. After a stereographic projection defined by $\{x_1, x_2, x_3\}/(1-x_4)$, the different values of angle $\phi$ describe a set of nested parallel tori in $\mathbb{R}^3$. The tori corresponding to, $\phi = 0$ and $\phi = \pi/2$ are great circles of $\mathbb{S}^3$ and in $\mathbb{R}^3$ correspond to the $C_\infty$ axes of the set of tori. For a given torus $\phi$, a torus knot T{$p,q$} can be expressed as,

$$\{R\cos(p\omega/q)\sin(\phi), R\sin(p\omega/q)\sin(\phi), R\cos(\omega)\cos(\phi)\}/(1-R\sin(\omega)\cos(\phi))$$

where, $p$ and $q$ are winding numbers describing the integer number of times the curve wraps around the two $C_\infty$ axes and $\omega \in [0, 2\pi pq)$. For all $\theta$ and $\phi$, T{1,1} forms the famous Hopf fibration [51], a fibered space where linked circles fill $\mathbb{R}^3$. Other T{$p,q$} knots form Seifert fibrations with different twist properties. Considering the LC director field, the requirement for winding numbers to be integer-valued is relaxed, in other words, T{$p,q$} can have irrational $p/q$ values within a director structure, so that different torus knots can simultaneously exist.

Through numerical simulations (Figs. 2 and 3), one can analyze the 3D director field configurations of torons in terms of the above formalism. For this, a series of streamlines are constructed by taking small spatial steps tangent to **n(r)** to form 3D curves in $\mathbb{R}^3$. Some streamlines terminate on confining surfaces or edges of the computational volume while others loop back on themselves to form closed loops such as the T{$p,q$} knots (Fig. 3). Furthermore, the allowed continuously-varying irrational $p/q$ values often result in streamlines that wind around the torus axes without ever completing a closed curve, which can be characterized by measuring the length of the streamlines (Fig. 3). These findings reveal that the toron configuration has spatially-varying director distortions deviating from the idealized 3D twisted structure that one could obtain by the stereographic projection, so that both the rate of the twist and T{$p,q$} depend on $\phi$ and the distance from the circular axis. This is because the toron combines the favorable 3D twisted region with some bend and splay distortions that aid in embedding the twisted director configuration in uniform far field while minimizing the overall free energy. Within the toron, we find that the smallest $\phi$ that yields a T{$p,q$} knot tangent to **n(r)** results in the Hopf link T{1,1}. Increasing $\phi$ results in different $p/q$ values, where some of the torus knots with small integer winding numbers that we find include the trefoil T{3,2}, pentafoil T{5,3}, and the quatrefoil T{3,4} torus knots (Fig. 3e-g). By simulating the response of the LC to applied electric field, we uncover how this electric field morphs the toron (Fig. 3a-d) and follow the evolution of these torus knots. As expected, the different torus knots on torus surfaces of constant $\phi$ never pass through each other as the structure morphs (Fig. 3e-g), but the contour lengths of the closed-loop knots tend to increase with voltage for sub-1 V applied voltages (Fig. 3e). This behavior of the streamlines tangent to **n(r)** that loop around both $C_\infty$ axes is dominated by energetically-favorable twist within an axially-symmetric toron structure at small voltages (Fig 3a,b,e-g). However, as the applied voltage is increased further and the soliton becomes asymmetric, few or none of the simple closed-loop streamlines can be found (Fig. 3c,d). The temporal evolution of streamlines is different upon turning voltage on and off (Video 1), so that the overall kinetic pathway of evolution of **n(r)** is non-reciprocal. Moreover, when the applied voltage is modulated, this dramatic non-reciprocal evolution of **n(r)** and the corresponding streamlines (Videos 1 and 2) takes place within each modulation period $T_m$, which, as we will see below, results in the translational motion of these solitons.

**III.B. Structure and topology of solitons stretched orthogonally to the far-field director.** Using laser tweezers, we can displace the top and bottom point defects of the toron laterally (say, along the y-direction, Fig. 4). Consequently, the skyrmion connecting the point defects becomes extended laterally in a direction perpendicular to the vertical far-field director at no applied fields (Fig. 4). The topology of this extended elementary skyrmion remains unchanged (e.g., mapping the vectorized **n(r)** from the cross-section orthogonal to the length of this extended skyrmion still covers the $\mathbb{S}^2$ once), however, the stretched field configuration lacks axial symmetry even without applied voltage. Similar to the unstretched skyrmion within the toron, which spans between substrates across the cell thickness, the laterally stretched configuration also terminates at two point singularities. These singularities appear as the points at which different colors meet within the Pontryagin-Thom construction (Fig. 4a). The Pontryagin-Thom construction here represents an isosurface of the zero z-component of **n(r)** (corresponding to the polar angle $\theta = \pi/2$), colored by the azimuthal orientation of the in-plane director. The vectorial presentation of **n(r)** reveal how the twist-embedding soliton is matched to the uniform background with the help of the same self-compensating hyperbolic point defects as within the toron (Fig. 4b-j), though these defects are now displaced laterally. The laser-assisted transformation of the toron into a stretched skyrmionic configuration resembles an inverse process of cholesteric fingers of the second type collapsing into a toron. In fact, the stretched skyrmionic configuration is nothing else but a cholesteric finger of the second type, studied previously [52], and is often found as a metastable configuration occurring spontaneously, without the assistance of laser tweezers.

Upon applying voltage to the cell with negative-$\Delta\varepsilon$ LC (Fig. 5), we again observe asymmetric morphing of the stretched skyrmionic structure. At applied fields above the well-defined voltage threshold [20], it becomes embedded in TIC (like in the case of a toron, Fig. 2). However, the topological hedgehog charges of the self-compensating hyperbolic point defects and the skyrmion number of the elementary skyrmion corresponding to the cross-section of the cholesteric finger of second type remain unchanged. As in the case of toron, the north-pole preimage which initially corresponds to the uniform far-field background at $U = 0$ V shrinks into an umbilical region, though this region itself is stretched along the length of the finger rather than being localized (Fig. 5) and embeds the extended skyrmion into TIC. In fact, one can think about the cholesteric finger at applied voltage as the corresponding stretched toron at the same applied voltage. Overall, our studies of skyrmionic field configurations show that the dramatic changes of **n(r)**, as in response to applied voltages and stretching by laser tweezers, can preserve topological characteristics and remain stable due to their nontrivial topology.

**III.C. Topology-preserving dynamics of switching between different states of solitons.** What are the kinetic pathways for the transformations between various topology-preserving states at different applied fields, like the ones shown in Fig. 2? The spatial locations of north-pole and south-pole preimages, as well as all other preimages, vary over time after turning voltage on and off while our topology-protected structures of solitons evolve towards equilibrium over time. During the initial response to switching the electric field on, the reorientation of initially-homeotropic background to spatially-varying **n(r)** is relatively fast, on the scale of tens of milliseconds that are typical for switching the LC director in cells of similar thickness [32]. This is then followed by a much slower process (tens of seconds) associated with further minimization of elastic free energy that re-defines the spatial configurations of the soliton embedded in TIC to correspond to the free-energy minimum at the corresponding voltage (Fig. 6a,b). At the initial stages, the localized preimages of the north and south poles appear at random locations with respect to each other, so that the south-north preimage dipole vectors have different orientations (Fig. 6a,b). These orientations of preimage vectors are initially determined by the spontaneous spatially-varying tilt orientations of the director in response to the applied field, as well as by presence of the additional self-compensating umbilical defects that eventually annihilate. With time, driven by the minimization of free energy of the background-TIC texture, these preimage vectors orientations synchronize to all point along the roughly same orientation (Fig. 6b), perpendicular to the uniform far-field tilt direction in the midplane of the TIC. This complex dynamic process involves motions of preimages of different $\mathbb{S}^2$-points relative to each other, (e.g. the motion of north-pole and south-pole preimages towards each other until they reside at a certain equilibrium distance). This process can be studied in a controlled way (Fig. 6c-g) by using holographic laser tweezers to accurately manipulate the localized regions corresponding to preimages of the north and south poles of the skyrmionic structure, as well as by using video microscopy to track their motion.

Upon application of voltage to a cell with elementary torons, POM frames of the localization of the north-pole preimage and formation of uniform TIC can be observed (Fig. 6a,c). With laser tweezers, we pull the north-pole and south-pole preimages apart and then release them to move freely towards each other at a constant applied voltage within the uniform surrounding TIC (Fig. 6c-e). When released, the preimages of the diametrically-opposite poles interact elastically at distances greater than 60 μm, eventually coming to rest at an equilibrium separation close to 3 μm. We observe this process at various voltages, ranging in amplitude within $U = 3.5 - 4.1$ V, with 1 kHz square waveform and then obtain the position versus time data shown in Fig. 6f,g by averaging over 10 videos. From the analysis of the spatial locations of the preimages within the different equally-spaced frames of optical microscopy videos, we also obtain velocities of preimages of the north and south poles of $\mathbb{S}^2$ as they approach and self-assemble within the localized solitonic structures (Fig. 7). A surface plot shown in Fig. 7a illustrates how these velocities behave as a function of the relative separation between the preimages of the poles and applied voltage. Interestingly, the south-pole preimage moves significantly faster than that of the north pole for all separations and voltages (Fig. 7a), which resembles the differences in velocities of motion of umbilical defects studied previously [53]. To analyze this physical behavior further, at the minimum and maximum applied voltages we take the derivative of the position versus time curve (Fig. 7b), thus obtaining the velocity versus time (Fig. 7c) and voltage dependence of the north- and south-pole preimage velocities. Log-log plots of the velocity and separation data in the range between 60 μm and 10 μm as the preimages are coming together, for the minimum (Fig. 7d) and maximum (Fig. 7e) applied voltages, also provide insights into the dynamics of evolution of the skyrmionic field configurations in applied field. The difference in slopes of the two plots further highlights the voltage dependence of the motion of preimages (Fig. 7d,e). The dependence of preimage velocities on the distance from each other is also apparent: as the preimages of the poles get closer, they move faster (Fig. 7). The speed anisotropy, or difference between the south-pole and north-pole preimage velocities, has a similar dependence on the separation distance (Fig. 8). We again see that the speed anisotropy depends on both applied voltage and separation distance (Fig. 8a). The maximum speed anisotropy increases with increasing the applied voltage (Fig. 8b), whereas the separation of the north- and south-pole preimages as measured at the maximum speed anisotropy peaks for each applied voltage decreases with increasing voltage (Fig. 8c). The symmetry-breaking dynamics of evolution of these skyrmionic structures embedded in a distorted far-field TIC may provide insights for understanding the physical mechanisms by which one can induce motion of the skyrmion with modulated voltage, as discussed below.

**III.D. Directional motion of extended skyrmionic structures.** An interesting observation is that all skyrmionic structures described above tend to shift laterally when the applied voltage is turned on and off. A highly controllable electrically-driven motion of the skyrmionic structures can be achieved in chiral nematic hosts with both positive and negative $\Delta\varepsilon$ by modulating the electric carrier signal at a lower modulation frequency [20]. The directions of soliton motion are in the plane of the LC cell orthogonal to the applied field **E** and depend on both the symmetry and kinetics of the director-field realignment in response to the time-varying amplitude of the modulated voltage. For example, we observe the electrically-driven motion (Fig. 9) of the extended skyrmionic structure stretched orthogonally to $\mathbf{n}_0$ at no fields, for which we have explored the energy-minimizing static field configurations at various applied voltages (Figs. 4 and 5). To characterize this motion, we use cholesteric fingers of the second type spanning through the entire width of the sample or, alternatively, we pin each end of the laser-generated structure at the locations of hyperbolic point defects using laser tweezers with a relatively high-power beam (100-200 mW). The latter method allows us to modify local surface boundary conditions at the cholesteric finger ends so that they become effectively adhered to the confining surfaces. In the chiral nematic LC with positive $\Delta\varepsilon$ (based on E7, see Table 1), we analyze both the translational motion and periodic changes of width of the solitonic structure in response to the modulated applied voltage (Fig. 9). When applying a square waveform modulation (Fig. 9a), we find that the width of the structure periodically oscillates in accord with the modulated voltage: the width, $w$, decreases dramatically when the $U = 2$ V potential is applied and slowly grows back to nearly its original value when $U$ is turned off. This periodic oscillation repeats within each modulation period and leads to a net directional translation of the structure over time (Fig. 9b). The net translation of the soliton arises from the fore-aft asymmetry of the shift corresponding to turning voltage on and off within each modulation period $T_m$. To gain further insights into the physical underpinnings of the observed voltage-driven dynamics of solitons, we analyze their periodic oscillations via probing the intensity profiles of the textures obtained using polarizing optical video microscopy (Fig. 9c,d). During each voltage modulation period, the width of the structure shrinks and grows back (Fig. 9e) while the structure translates by tens and hundreds of microns in the x-direction orthogonal to the skyrmionic structure (Fig. 9f), with the structure exhibiting similar cross-sectional intensity profiles at initial and final positions (Fig. 9g).

We also demonstrate translational motion of an extended skyrmionic structure in a negative $\Delta\varepsilon$ material (based on the MLC-6609 nematic host, Table 1) while using a similar square-wave modulation voltage driving scheme (Fig. 9a). An extended skyrmionic structure, which is a fragment of the cholesteric finger of the second type, is pinned at its ends to observe the entire length of the structure using video microscopy (Fig. 10). Upon applying the modulated electric field, we see that a dark region (the preimage of the north pole in applied field) appears on the right side of the structure to embed it in the far-field background. The skyrmionic structure translates along a vector connecting this north-pole preimage to the south-pole preimage which corresponds to the dark region within the finger that spans the length of the structure in the polarizing optical micrograph. When the voltage modulation is on, the end-pinned skyrmionic configuration adopts an archlike bent profile, which straightens and translates back to its equilibrium position and conformation when the field is turned off or in applied field at carrier frequency without the amplitude modulation. We induce this type of dynamics repeatedly, hundreds of times, with the full relaxation of stretched skyrmionic structure to its unbent state taking several minutes (Fig. 10). This slow relaxation of the relatively short finger fragment with pinned ends to a straight configuration is driven by the minimization of free energy (which can be estimated as a product of the finger's line tension and length) at no applied fields, but fingers spanning the entire sample width can move much farther, similar to the example shown in Fig. 9. Since the electrically-driven translation of skyrmions occurs for various configuration geometries and in host materials with both positive and negative dielectric anisotropy, these observations support our hypothesis that the motion arises from the asymmetric, non-reciprocal rotation of the director field during the on and off fragments of the voltage modulation period [20].

**III.E. Field-driven translation of torons.** Characterization of directional motion of skyrmionic structures shaped as torons is presented in Fig. 11. The experimentally-observed asymmetric spatial shift of the toron in response to turning voltage on and off (Fig. 11a) is reproduced by numerical modeling using the RBF-FD relaxation method (Fig. 11b, Video 1) and is similar to that we characterized for the cholesteric fingers of the second type (Fig. 9). This asymmetry of the lateral shift originates from the different balances of torques when the applied field is on and off, which can be used to controllably translate torons (Fig. 11c,d). The control parameters of the used voltage driving scheme (insets of Fig. 11c,d) include the low and high voltage amplitudes $U_1$ and $U_2$, frequency of modulation $f_m$, and the duty cycle of the modulation, all allowing for the control of velocity and direction of the toron motion. By systematically scanning parameters $U_1/U_2$ and $f_m$, we observe that the direction of motion can be controllably reversed (Fig. 11c,d). For example, starting at a relatively low modulation frequency ($f_m$

= 2 Hz) and 70% duty cycle, we observe a positive velocity for fill ratios $0.3 < U_1/U_2 < 0.6$ and negative velocity for fill ratios $U_1/U_2 > 0.6$ (Fig. 11c), where positive displacements and velocities are defined along the x-axis pointing in the direction from the north-pole preimage to the south-pole preimage (Fig. 11a inset). Similarly, but now for the constant fill ratio $U_1/U_2 = 0$, 75% duty cycle and while varying $f_m$, we see positive velocities at modulation frequencies below 4.5 Hz and negative velocities above 4.5 Hz (Fig. 11d). To provide further insights into the physical underpinnings behind this controllable directional motion, we construct the temporal evolution of streamlines of the director field that take place within each $T_m$, highlighting its non-reciprocal nature (see the example in the Video 2 for low modulation frequency corresponding to positive velocities). During this motion, the periodically-repeated asymmetric squeezing of the skyrmionic field configuration induced by turning the voltage on and off within each $T_m$ results in a squirming motion of the soliton. The directionality and velocity amplitude of this motion can be controlled based on tuning the characteristic times of oscillation of the modulated voltage relative to the response times of director reorientation in response to turning the field on and off. These findings show how two geometrically different skyrmionic structures with similar topology, the elementary toron and the cholesteric finger of the second type, share not only the value of the skyrmion number but also largely similar response to the oscillating electric field: both types of solitons move in directions orthogonal to the applied field and, depending on the parameters of the voltage driving scheme, move parallel or anti-parallel to the vector connecting the north-pole preimage to the south-pole preimage.

**III.F. Tracer nanoparticles and analysis of director reorientation-driven backflows.**

To probe the possible role of backflows and other types of flows in the field-driven motion of topological solitons, we use metal and semiconductor nanoparticles as tracers. The rodlike plasmonic nanoparticles, with the surface plasmon resonance spectra featuring transverse and longitudinal peaks (Fig. 12a), allow us to track their spatial translations based on light scattering when imaged in the dark-field microscopy mode (Fig. 12b,c). Using particle tracking and video microscopy, we can indeed detect weak flows in the studied cells and characterize the anisotropic and depth-dependent nature of these flows (Fig. 12b,c), though we find that these weak flows cannot account for the translational motion of solitons (Fig. 12d-g). The spatial displacement amplitude corresponding to each period of voltage modulation $T_m$ increases with the voltage amplitude, becoming direction-dependent at higher fields, but it remains relatively small within the range of used voltages. Because of the presence of director twist within the TIC, the depth dependence of the displacement amplitude is asymmetric with respect to the cell midplane (Fig. 12c), with only very small displacements observed close to the confining substrates, as expected. Imaging of both the plasmonic nanorods based on scattering (inset of Fig. 12c) and semiconductor nanocubes based on luminescence (Fig. 12d,f) shows that the solitons can move past the multiple tracer nanoparticles dispersed within the cell bulk, close to the midplane of the LC cell (Fig. 12d,g). These findings support the notion that the soliton translation does not necessarily have to involve mass transport [20]. The rotational dynamics of **n**(**r**) within these spatially-localized structures allows for advancement in different directions without relying on the actual LC fluid flows. This observation is also consistent with the fact that key features of our experimental findings can be reproduced numerically based on a model that involves only the rotational dynamics of the director, but no fluid flows (Fig. 11). We note, however, that both the semiconductor nanocubes and gold nanorods follow the motions of torons and cholesteric fingers when these particles get entrapped within the singular cores of point defects at the ends of skyrmionic tubes, as discussed in our earlier studies [54].

**III.G. Transport of cargo using solitons.** In addition to transporting the singular-defect-entrapped nanoparticles, as mentioned above [54], the dynamic evolution of skyrmionic textures that lacks time-reversal symmetry can also be used for transporting microparticles. We investigate the skyrmion's cargo-carrying abilities by embedding a melanin resin microparticle of 3 μm in diameter with planar degenerate surface anchoring between the two hyperbolic point defects (Fig. 2), thus entrapping our cargo within the solitonic configuration. The static 3D director structure of such torons entrapping particles with tangential boundary conditions has been reported earlier [55], so here we focus on their dynamic motion, which is then compared to that of torons without particles (Fig. 13). The microparticle embedded within the toron configuration can be observed under the optical microscope in a transmission mode without polarizers (Fig. 13a). When the combined motion of the toron with the microparticle cargo is compared side-by-side to the squirming motion of such a skyrmion without cargo (Fig. 13b), we find that the translational motion is not significantly hindered by the addition of cargo (Fig. 13c). Furthermore, similar to the case of such skyrmionic configurations without cargo (Fig. 11), we demonstrate directional control of the skyrmion motion with cargo by changing the modulation frequency (Fig. 13d). At higher modulation frequency, $f_m = 8$ Hz, the toron embedding cargo moves along the vector connecting the south-pole and north-pole preimages, and at lower modulation frequency of $f_m = 2$ Hz, the motion is reversed to point in the direction along the vector connecting the north-pole and south-pole preimages (Fig. 13e). These examples demonstrate the capability of a controllable transport of cargo along directions dependent on frequency of the applied electric field.

**III.H. Collective dynamics of active solitons.** Finally, we demonstrate collective motion of many solitons with voltage modulation (Fig. 14). By generating many skyrmions, we can selectively drive self-assembly of pairs, chains, and arrays of solitons within the far-field TIC by relying on the elasticity-mediated self-assembly described elsewhere [47]. Using an applied field of $U = 4.5$ V with the 1 kHz carrier signal modulated at $f_m = 1$ Hz, we then observe and compare motion of a self-assembled soliton pair and that of a single soliton (Fig.14a) while all trajectories of motion are guided to follow an arched path by controlling the far-field TIC. Under the same voltage-driving conditions, we also drive and observe collective motion of a chain of three solitons along a straight-line path (Fig.14b,c). To compare the motion of single solitons next to soliton pairs across a given region of our cholesteric cell, we set up a "race" where such solitons are initially self-assembled along a straight line and start moving at the same point in time, when voltage is modulated at $f_m = 2$ Hz (Fig.14d). We find that the single solitons move faster than the soliton pairs (Fig.14e). Additionally, we observe a similar difference in translational velocity for a race between a trio-chain, pair, and single solitons (Video 3), indicating that individual solitons are always faster and the bigger the chainlike assemblies of solitons, the slower they become. Interestingly, with modulation at $f_m = 1$ Hz the solitons tend to spread apart within the self-assembled chains as they move, so that the inter-soliton distances within chains by a factor of 2-3 when they move as compared to the stationary ones (Fig. 14a-b). However, this spread between solitons is no longer present with modulation at $f_m$ = 2 Hz (Fig. 14d,f). Therefore, we can tune both the directionality of chain motion and spreading distance between solitons within a chain. When we use elasticity-mediated self-assembly [47] to guide many solitons to self-assemble into a close-packed array and then modulate voltage at $f_m$ = 2 Hz, we observe dramatic transformations of kinetically re-assembled structures of solitons that can be guided to produce a long chain or many separate linear chains branching off from the initial array (Fig.14f). By tuning the applied voltage amplitude and $f_m$, as well as by controlling the far-field TIC, we can control the motion directions and reconfigurable kinetic assemblies of these solitons at will (Videos 3 and 4). As an example, we show movement and transformation of various self-assembled structures prompted by changing the modulation frequency from $f_m$ = 2 Hz to $f_m$ = 8 Hz (Video 4).

## IV. DISCUSSION AND CONCLUSIONS.

Although the different solitonic structures we study have very different appearance under a polarizing microscope, stretch in different directions with respect to the homeotropic $\mathbf{n}_0$ at no fields, and are often called elementary torons and cholesteric bubbles (Figs. 2,6-8,11-13) or cholesteric fingers of the second type (Figs. 4,5,9,10) [9,17,20,45,46,52, 56], they share the topology of an elementary skyrmion tube terminating on singular point defects. Indeed, the torons and fingers can be inter-transformed via optical manipulation using laser tweezers without changing their topology. The dynamics they exhibit are also similar, though the structural features and positive versus negative dielectric anisotropy can result in somewhat different selection of the motion directions (Figs. 9-14). What are the physical underpinnings of the studied dynamics of these solitons? To answer this question, it is instructive to recall some of the basics of LC display functioning.

A widely-known effect in physics of information displays is that the response of the LC director to turning the external electric field on is faster than in case of turning it off, with the characteristic rising and decay response times related as $\tau_{rising} = \tau_{decay}/[(U/U_{th})^2 - 1]$, where $U_{th}$ is a certain threshold voltage above which the director realignment starts because of the strength of electric torque overcoming the elastic torque [57]. At high voltages $U \gg U_{th}$, the difference between the rising and falling director response times can be several orders of magnitude and stems from the different torque balances present when voltage is turned on and off: the electric torque is balanced by elastic and viscous torques when the field is on, but only elastic and viscous torques are present and balanced upon turning voltage off [32,57]. The uniform and distorted configurations of the director field in the display pixels are typically translationally invariant, so that the non-reciprocal rotation manifests itself only through the translationally invariant changes of optical characteristics such as phase retardation and polarized light transmission (though various backflow effects may need to be accounted for at high voltages [57]). The situation is very different for the spatially-localized, topology-stabilized solitonic structures (Figs. 9-11 and 13), which translate in the lateral directions because of the non-reciprocal rotation of the constituent 3D **n**(**r**). In a pixel of a typical display, the applied electric field deforms the initially-uniform director and the elastic free energy accumulated within this deformation then drives the slower relaxation of the director back to the uniform state. The director structure of solitions is three-dimensional and spatially non-uniform, but the dielectric coupling of **n**(**r**) with **E** morphs it significantly, along with deforming the structure of the surrounding LC background (Figs. 2,4,5). The elastic free-energy cost associated with this deformation drives relaxation of the solitonic configuration back to the initial state that minimizes energy at no applied fields. The non-reciprocal nature of director rotation in response to switching voltage on and off yields the translational motion of solitons in the lateral directions. In principle, one could also expect soliton motions along the applied field direction in LC samples much thicker than the dimensions of solitons, but this is precluded in our case by the thin-cell confinement where solitons span through the entire cell gap. Within each period $T_m$ of modulated voltage, the topological solitons are squeezed due to the coupling of **n(r)** with **E** during the "field-on" cycle and relax to minimize the elastic free energy during the "field-off" cycle of $T_m$, where the non-reciprocity of the director rotation translates the localized solitonic structure laterally within the LC cell. This periodic morphing of the localized director field structure resembles squirming waves in biological systems [20, 25], albeit our chemically homogeneous solitons have no cell boundaries, density gradients or interfaces, so that the similarity is only at the level of the non-reciprocal dynamics. The squirminglike dynamics of localized director structures may also play a role in mediating translation of various colloidal particles around which the director is either asymmetric even at no fields, or the symmetry of director realignment is broken during switching, though this mechanism has not been considered in previous studies [58] and will need to be considered in future works.

The studied translation of topological solitons results from the conversion of electric energy into elastic energy and into motion. Similar to the cases of active colloidal particles and other active soft-matter systems [21], the energy conversion happens at the scale of individual particlelike solitons, though the oscillating energy-supplying field is applied globally to the entire sample typically $10^8$-times larger than the lateral area of the soliton. The directional motion of individual and multiple torons, which are axially symmetric and embedded in the far-field director orthogonal to cell substrates at no fields, results from the symmetry breaking during switching (Fig. 6a,b). The spontaneous selection of the motion directions happens as a result of the synchronization of the director tilt direction within the TIC (Fig. 6b), driven by elastic free energy minimization at an applied field. Once the midplane far-field director of TIC becomes uniform, without the additional umbilical defects, the vectors connecting the north-pole and south-pole preimages all point in the same direction and the resulting motion is either parallel or anti-parallel to these vectors, depending on the modulation frequency (Fig. 6b, 11-14). Unlike the torons, the laterally-stretched skyrmionic structures lack axial symmetry even at no fields, so that their motion direction is always orthogonal to their length (Figs. 9 and 10), but has similar origins otherwise. The synchronization of the motion direction that we observe for multiple solitons (Fig. 6a,b) also resembles synchronization of motion directions of defects in active matter recently studied by Dogic and colleagues [59]. The examples of collective out-of-equilibrium dynamics of solitons (Fig. 14) also bring about the resemblance with the behavior of both active colloidal particles and defects within active nematics [21-26]: the motion of each soliton influences that of their neighbors and kinetically self-assembled structures are significantly different from the ones obtained in the equilibrium. Our preliminary demonstrations (Figs. 13 and 14) show that the facile response of LCs, along with the diversity of means available for controlling **n(r)**, have a great potential to expand the scope and means of realizing guided motion of active colloidal particles with and without cargo [60,61].

Although the use of high carrier frequency in our study allows us to eliminate the possible role of ionic impurities in defining dynamics of the topological solitons, various electrophoretic and electrokinetic effects can be used to further enrich the possibilities for the control of the soliton dynamics [56]. Although the use of low voltages in our study precluded a significant role of backflows, one can potentially also design voltage-driving schemes where backflows could further accelerate motions of solitons or alter their motion directions, similar to how researchers can use backflow effects to speed up director switching in electro-optic LC devices [62]. On the other hand, the presented solitons with the topology of elementary skyrmions and structures that often locally resemble torus knots, Hopf, and Seifert fibrations may provide a model system for the study out-of-equilibrium phenomena related to field transformations within condensed matter and beyond. Finally, our study promises to bridge two so far separate research directions involving solitons, the study of solitary waves in fluid dynamics [1,2] and the research on topological solitons in field theories [3-20], where the interplay of different type of nonlinearities can yield rich fundamental behavior with out-of-equilibrium topological transformations and structure-and topology-preserving motion.

To conclude, we have demonstrated that the non-reciprocal response of LCs containing solitonic structures to external fields can allow for engineering spatial translation of solitons with and without carrying cargo, such as nanometer- and micrometer-sized particles. This directional electrically-driven motion can be exhibited by a host of topologically-protected structures, with different symmetry of localized director configurations and in LC materials with different types of response to external fields (e.g. LCs with positive and negative dielectric anisotropy). Although our non-annihilating solitons provide many advantages for experimental exploration of such topological dynamics, these studies can be extended to more conventional singular defects, where electric and magnetic driving schemes can be potentially designed to drive defects towards each other or apart, thus enabling the means of controlling out-of-equilibrium dynamics of soft matter with long-term-stable defects. Although we focused here on $\pi_2(\mathbb{S}^2)=\mathbb{Z}$ (and $\pi_2(\mathbb{S}^2/\mathbb{Z}^2)=\mathbb{Z}$ for the nonpolar **n(r)** ) solitons with the same terminating point defects, these studies can be extended to the $\pi_2(\mathbb{S}^2)=\mathbb{Z}$ (and $\pi_2(\mathbb{S}^2/\mathbb{Z}^2)=\mathbb{Z}$ for the nonpolar **n(r)** ) Hopf solitons discovered recently in similar confined chiral nematic LC systems, which will be pursued in our future studies.

**ACKNOWLEDGEMENTS.**

We acknowledge discussions with A. Bogdanov, N. Clark, O. Lavrentovich, P. Pieranski, J. Selinger, M. Tasinkevych, J.-S. Tai, S. Thampi, J. Yeomans and S. Zumer. We thank H. Mundoor and S. Park for providing colloidal nanocubes that we use as one type of tracer particles, as well as we acknowledge technical assistance of T. Lee, Q. Liu, H. Mundoor, S. Park and B. Senyuk. This research was supported by the National Science Foundation Grant DMR-1410735 (I.I.S., P.J.A., T.J.B., and H.R.O.S.) and NSF Graduate Research Fellowship Program Grant DGE 1144083 (H.R.O.S.). This work utilized the RMACC Summit supercomputer, which is supported by the National Science Foundation (awards ACI-1532235 and ACI-1532236), as well as the facilities of the Materials Science and Engineering Center at CU-Boulder (National Science Foundation Grant DMR-1420736).

**References.**

**Figures:**

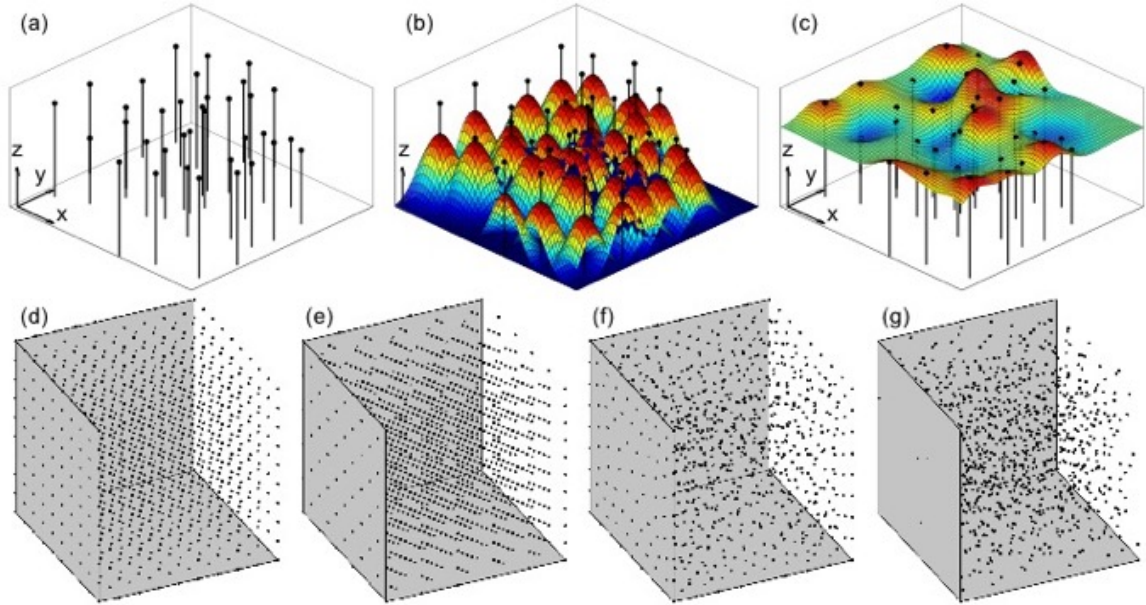

FIG. 1. RBF method for defining nodes. (a) Random points scattered in two dimensions. (b) A Gaussian basis set illustrated by centered Gaussian surfaces at the location of each node in (a). Colors represent the relative height of the surfaces. (c) A surface that passes through the points constructed from a unique linear combination of the Gaussians from (b). (d) Periodic square lattice, (e) face-centered cubic lattice, (f) tetrahedral mesh, and (g) Halton-type scattered node distributions in 3D.

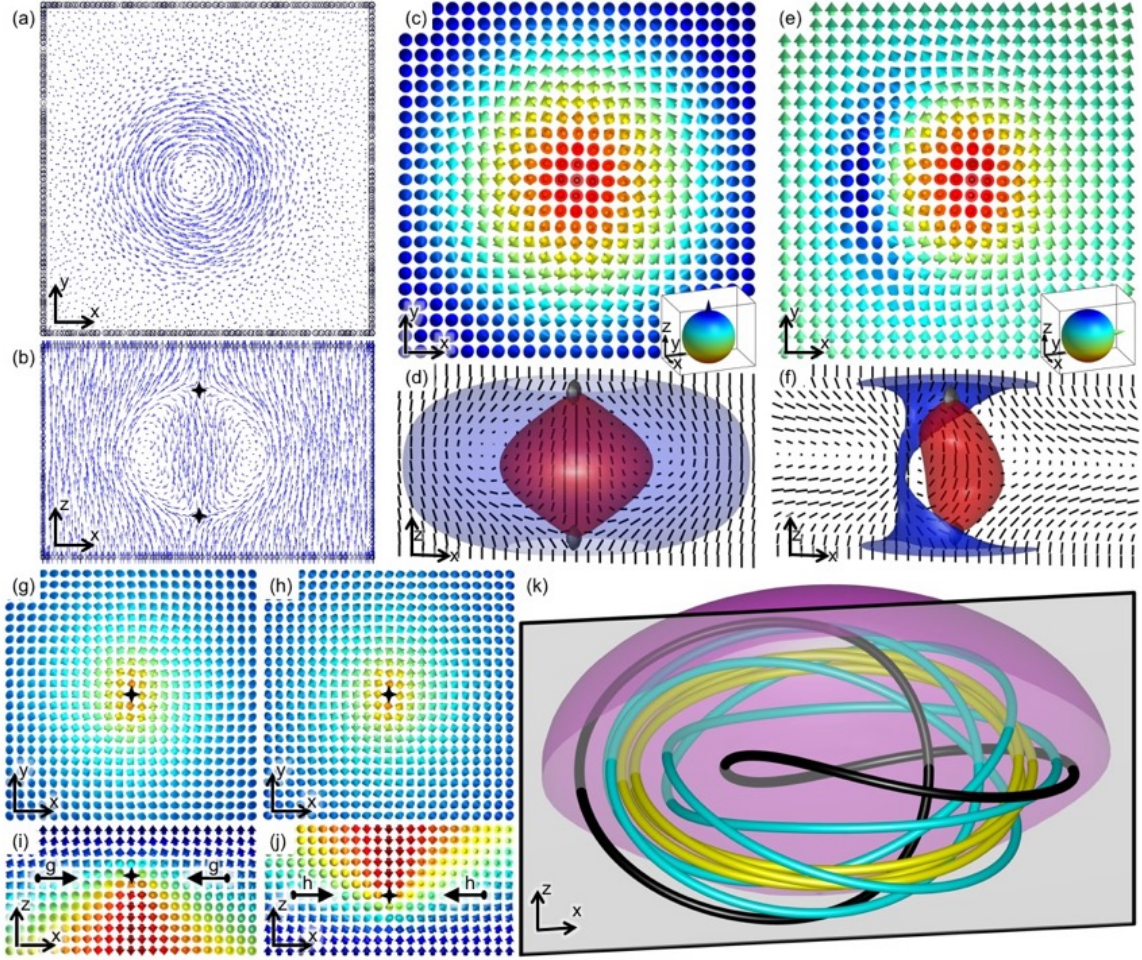

FIG. 2. Structure of a toron with the topology of an elementary skyrmion and knotted streamlines. A skyrmion embedded in a confined chiral nematic LC with (a-d, g-k) no applied field, $U = 0$ V, with uniform far-field director $\mathbf{n}_0 = \{0,0,1\}$ and (e-f) applied field, $U = 3.1$ V, with uniform far-field director $\mathbf{n}_0 = \{0,1,0\}$. The direct comparison shows agreement of RBF-FD simulated 2D cross-sectional midplanes in (a) x-y and (b) x-z of $\mathbf{n}(\mathbf{r})$ with the (c) traditional FD simulated 2D x-y cross-sectional midplane of $\mathbf{n}(\mathbf{r})$ using arrows colored according to the corresponding points on $\mathbb{S}^2$ (inset). The orientation of $\mathbf{n}_0$ on $\mathbb{S}^2$ is denoted using cones. (d) Corresponding 3D representation of $\mathbf{n}(\mathbf{r})$ described in part (c) by means of isosurfaces representing the z-component of the director, $n_z = 0.95$ (blue), $n_z = -0.2$ (red), regions around singularities (gray) and an x-z cross-section of the director through both point defects visualized using black line segments. (e) 2D x-y cross-sectional midplane of a vectorized skyrmion with applied field. (f) Corresponding 3D representation of $\mathbf{n}(\mathbf{r})$ described in part (e) by means of isosurfaces described in (c). A zoomed-in view at the x-y cross-section through point defects of skyrmions, confined in a chiral nematic LC with no applied field, (g) above and (h) below the skyrmion tube and x-z cross-section through point defects (i) above and (j) below the skyrmion. Additional arrows in (i,j) denote locations of (g,h) cross-sections. (k) Streamlines tangent to director within the twisted region of a toron (magenta surface) that trace a Hopf link (black), pentafoil knot (cyan), and quatrefoil knot (yellow). Chiral mixture is based on MLC-6609 and ZLI-811.

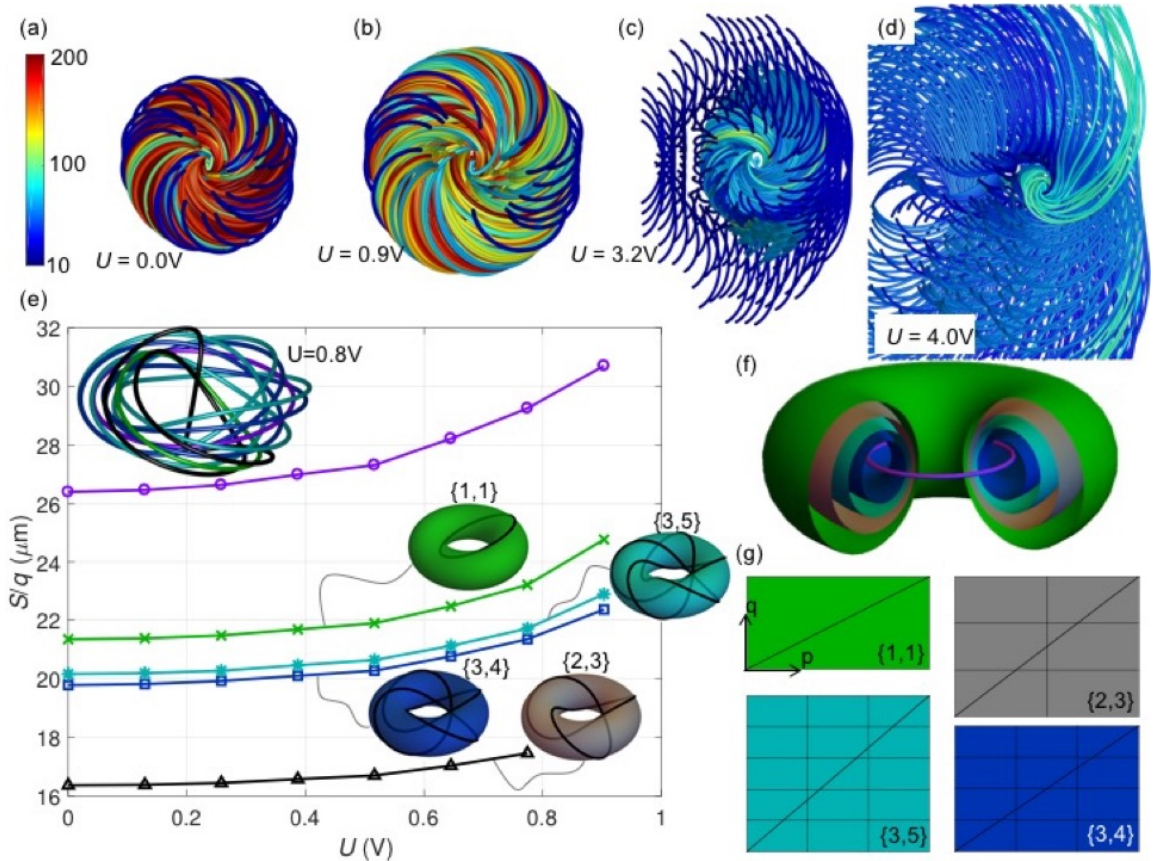

FIG. 3. Torus knots and electric switching of director streamlines. (a-d) Streamlines tangent to the director field lines at various applied voltages. The streamlines originate from the cell midplane and propagate along the director field until they terminate on a substrate, close on themselves to form knots or become longer than 200 μm. The color bar indicates lengths of continuous streamlines, with blue representing the short streamlines that crash to the substrates quickly (~10 μm) and red streamlines loop around the torus many times and often form a $\{p,q\}$ torus knots ($\geq$ 200 μm). (e) Contour length, $S$, over the winding number, $q$, of some of the torus knots that are found for $U < 1$ V, plotted as a function of applied voltage. The top left inset shows examples of the tracked torus knots and unknots at $U = 0.8$ V. The top purple curve (circles) is the circumference of the $\phi = \pi/2$ $C_\infty$ axis. Crosses, asterisks, squares, and triangles mark length of the Hopf unknot, pentafoil knot, quatrefoil knot, and trefoil knot, respectively. The torus knots tracked each have insets representing there topology and $\{p,q\}$ winding numbers, with the corresponding voltage dependencies of the contour lengths indicated. (f) Representative torus surfaces between $\phi = \pi/4$ and $\phi = \pi/2$ near to where the respective torus knots and unknots are found including the $\phi = \pi/2$ $C_\infty$ axis. (g) Rectangles schematically representing the unwrapped torus surfaces shown in (f) with the same colors, where each torus is shown $p \cdot q$ times to indicate how many times and how the director streamline slides on the torus surface before looping on itself. The thin black lines indicate the director streamlines that loop around the two axes of the torus to form various torus knots. Chiral mixture is based on MLC-6609 and ZLI-811.

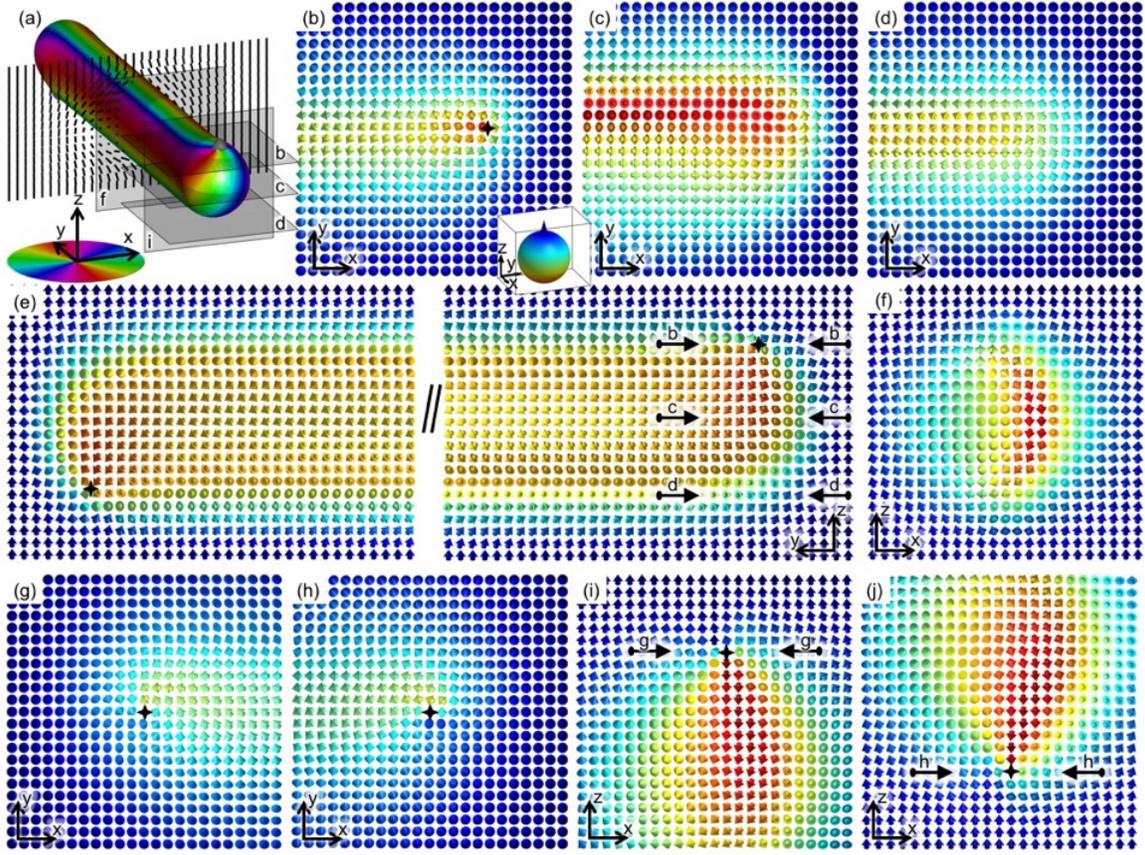

FIG. 4. Structure of a laterally-stretched skyrmion at no fields. (a) 3D and (b-j) 2D representations of the stretched skyrmion structure embedded within the uniform far-field director $\mathbf{n}_0 = \{0,0,1\}$. (a) The 3D structure is visualized using the Pontryagin-Thom construction, small circular gray isosurfaces representing regions around singularities, and an x-z cross-section of the director through the skyrmion visualized using black line segments. 2D x-y cross-sectional planes through the (b) top (c) midplane and (d) bottom of vectorized $\mathbf{n}(\mathbf{r})$ using arrows colored according to the corresponding points on $\mathbb{S}^2$ (inset) and the orientation of $\mathbf{n}_0$ on $\mathbb{S}^2$ denoted using a cone. The black stars represent the point singularities (corresponding to the gray surfaces in part a). (e) A y-z cross-section along the length of the structure. (f) An x-z cross-section through the midplane of the structure. Additionally, the x-y cross-sections through point defects (denoted with black stars) of the structure (g) near the top substrate and (h) near the bottom substrate and x-z cross-section through point defects (i) near the top and (j) near the bottom of the structure. Chiral mixture is based on MLC-6609 and ZLI-811.
48...

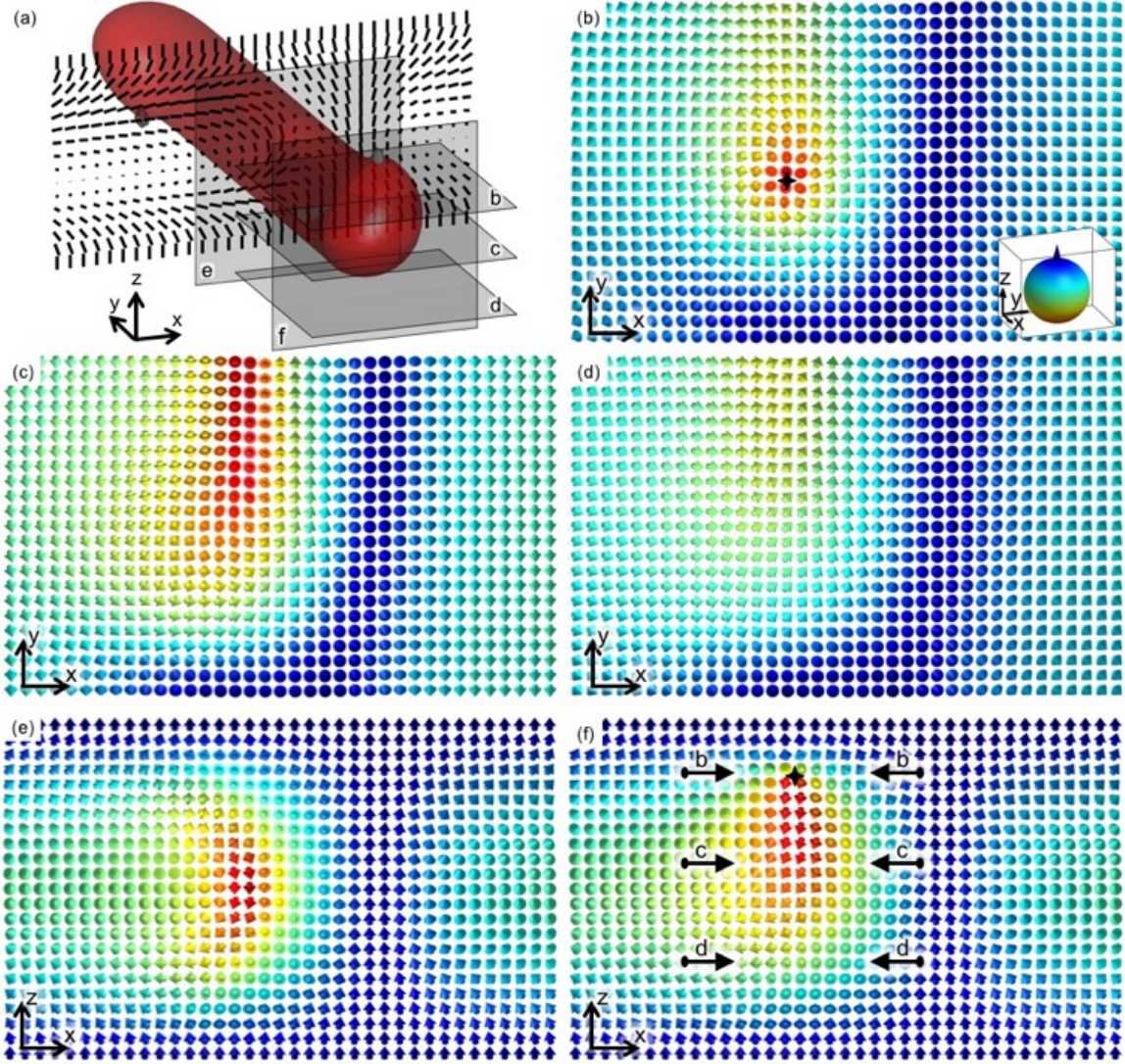

FIG. 5. **Structure of a laterally-stretched skyrmion in an applied field.** (a) 3D and (b-f) 2D representations of a stretched skyrmion structure at $U = 4$ V. (a) Isosurfaces representing the z-component of the director $n_z = -0.2$ (red) and regions around singularities (circular gray surface) and x-z cross-section of the director through the skyrmion visualized using black line segments. 2D x-y cross-sectional planes passing through the (b) top (c) midplane and (d) bottom of vectorized **n(r)**-configurations using arrows colored according to the corresponding points on $\mathbb{S}^2$ (inset) and the orientation of $\mathbf{n}_0$ on $\mathbb{S}^2$ denoted using a cone. The black stars represent the point singularities (corresponding to the gray surfaces in part a). 2D x-z cross-sections of the extended structure through (e) the skyrmion and (f) one of the hyperbolic point defects near the top substrate. Additional arrows in (f) denote locations of the (b,c,d) cross-sections. Chiral mixture is based on MLC-6609 and ZLI-811.

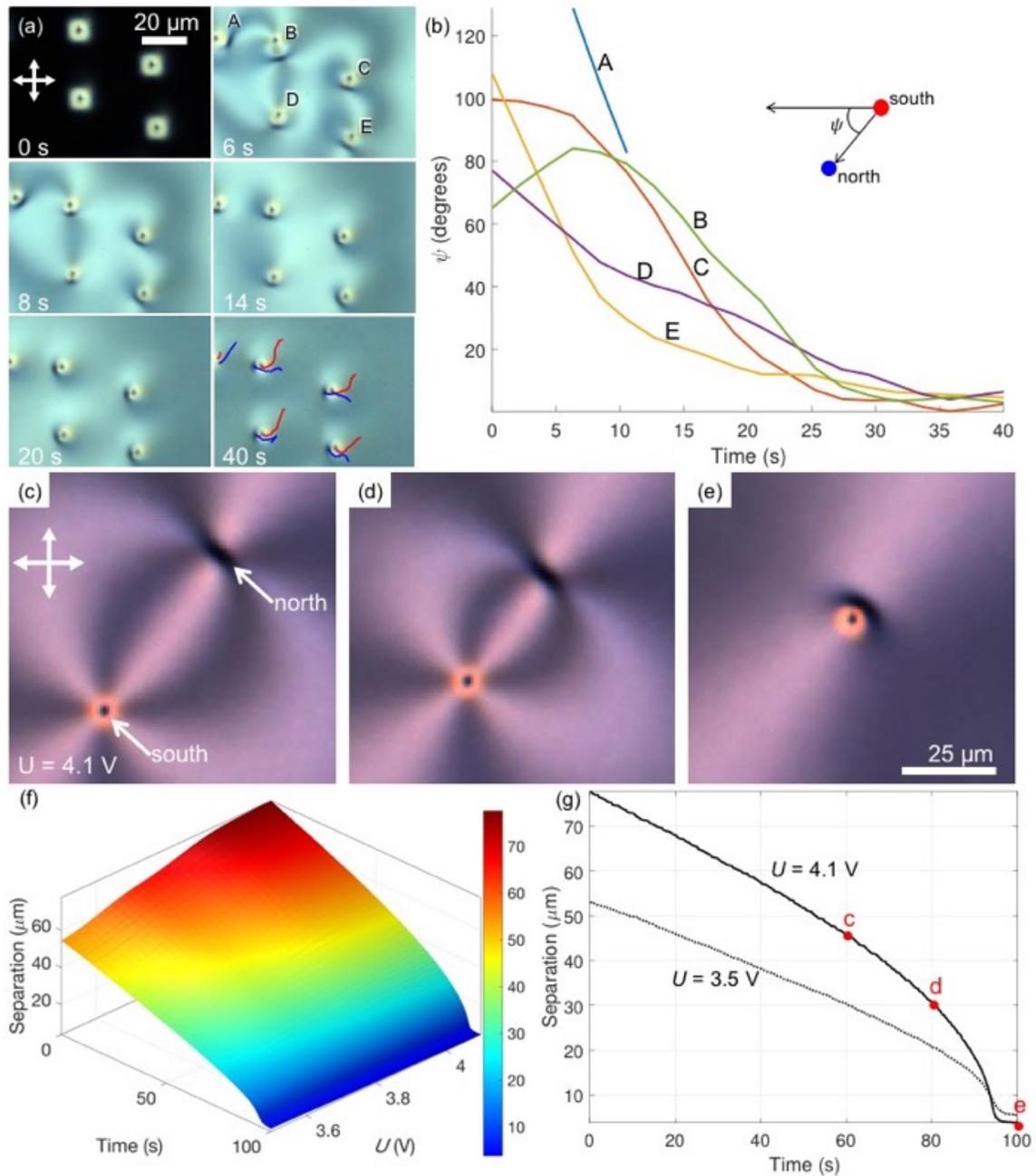

FIG. 6. Formation of skyrmions and synchronization of south-pole to north-pole preimage vectors. (a) POM images showing gradual self-alignment of preimage vectors within the TIC. Trajectories of south-pole (red) and north-pole (blue) preimages after applying voltage are overlaid on the last frame. The elapsed time is indicated on the POM video frames. (b) An angle characterizing the azimuthal orientations of the preimage dipoles versus time for the five solitons labeled in (a). (c-e) POM images showing interaction of the north-pole and south-pole preimages after they were pulled apart by laser tweezers. Crossed polarizers in POM images are marked with white double arrows. North-pole and south-pole preimage separation (f) for many applied voltages, represented by a surface colored according to the magnitude of separation, and (g) for the minimum and maximum voltages applied. The points corresponding to the images in (c-e) are marked by red dots on the $U = 4.1$ V curve. Chiral mixture is based on MLC-6609 and ZLI-811 where thickness and pitch are about 10 μm.

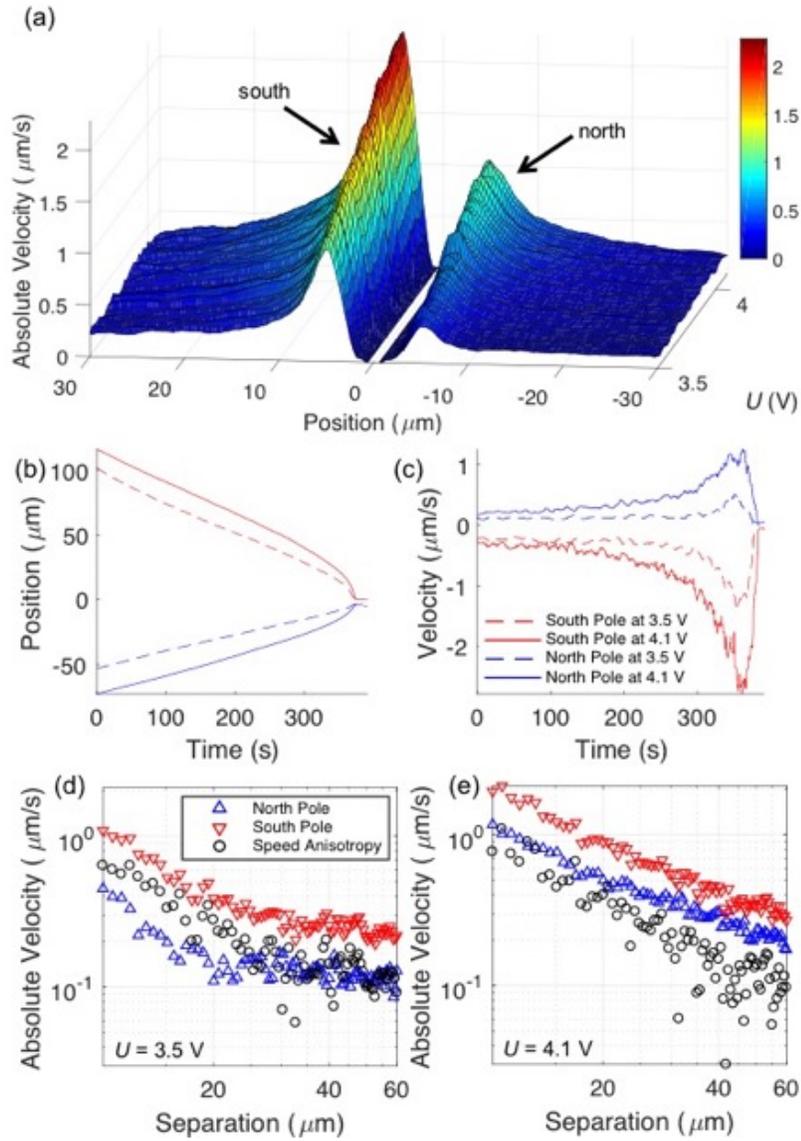

FIG. 7. Dynamics of skyrmion's preimages. (a) North-pole and south-pole preimage velocities during skyrmion formation, where the surface is colored according to the velocity magnitude. (b) Position and (c) velocity of the preimages during attraction for the two extreme voltages. (d, e) Logarithmic representations of the preimage velocities and the velocity speed anisotropy during attraction at (d) $U = 3.5$ V and (e) $U = 4.1$ V. Chiral mixture is MLC-6609 and ZLI-811 where thickness and pitch are about 10 μm.

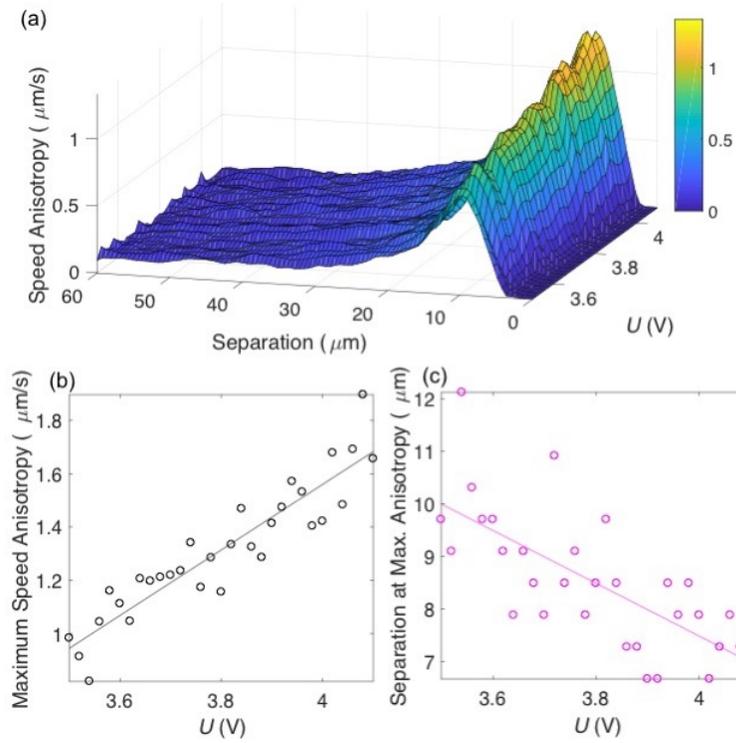

FIG. 8. Anisotropy of north-pole and south-pole preimage velocities. (a) Speed anisotropies of the preimages, where the surface is colored according to magnitude of speed anisotropy. (b, c) Voltage dependence of (b) the maximum speed anisotropy and (c) the north-pole and south-pole preimage separation measured at the maximum speed anisotropy. Chiral mixture is MLC-6609 and ZLI-811 where thickness and pitch are about 10 μm.

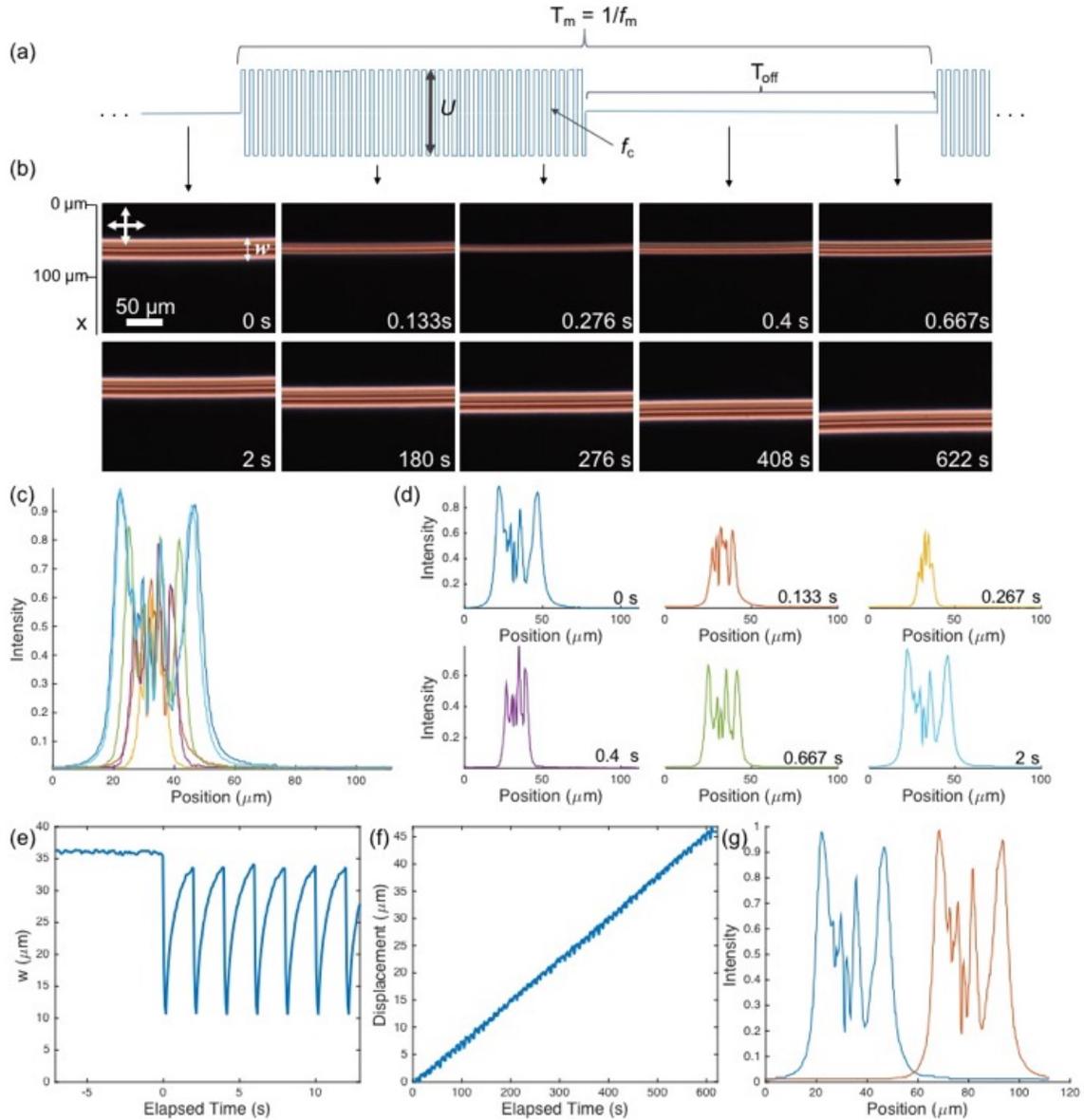

FIG. 9. Translational motion of a stretched skyrmion structure. (a) Voltage profile of the square waveform used for modulation, where $f_m = 0.5$ Hz. The black arrows indicate (b) corresponding POM images of a stretched skyrmion structure in the sample (top). Translational motion of the structure is shown over many modulation cycles (bottom). White double arrows indicate the polarizer orientation. (c, d) Intensity profiles of the extended structure within one period of voltage modulation, extracted from POM images shown in (b). (e) Periodic change in width, $w$, as defined in (b), of the structure, where voltage modulation starts at 0 s. (f) Lateral displacement of the structure over time. (g) Intensity profiles corresponding to the initial (blue) and final (orange) positions of the structure extracted from the corresponding POM images. Chiral mixture is E7 and CB-15 where thickness and pitch are about 30 μm.

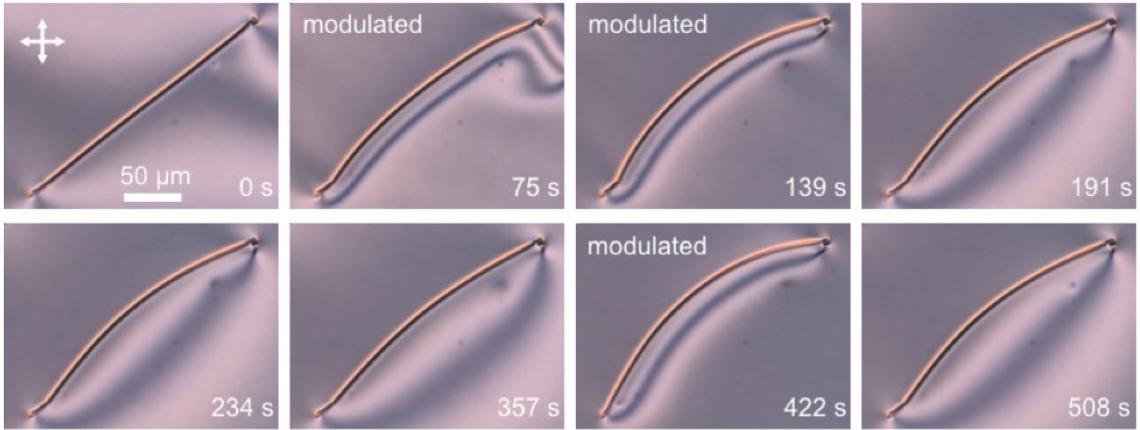

FIG. 10. Translational motion of a stretched skyrmion structure. The thickness and pitch are close to 10 μm. The applied voltage is always $U = 4.2$ V, which is modulated at $f_m = 2$ Hz where noted in top left corner and not modulated otherwise. The polarizer orientations are marked with white double arrows and elapsed time is noted in the bottom right corner corners of images. Chiral mixture is MLC-6609 and ZLI-811 where thickness and pitch are about 10 μm.

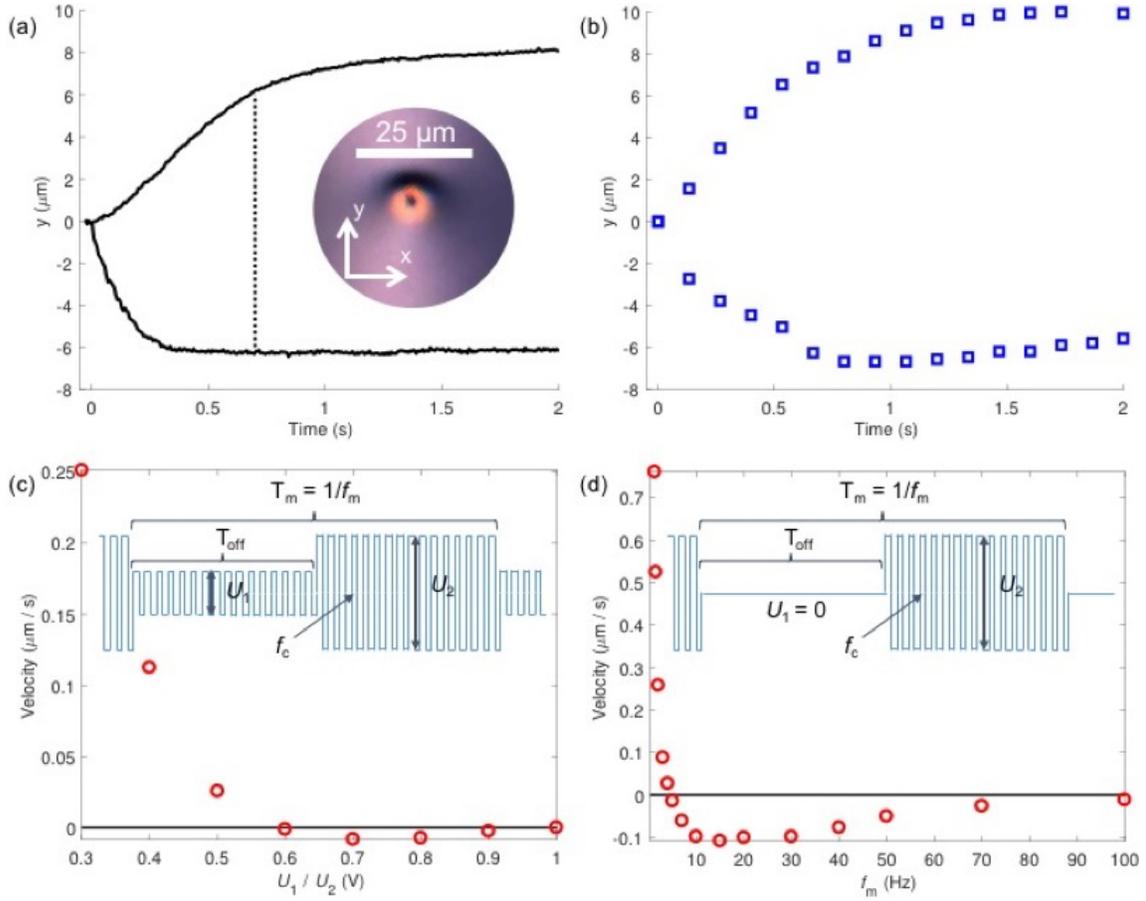

FIG. 11. Directional motion of torons. (a) Experimental and (b) RBF-FD simulated skyrmion position when voltage modulation is turned on (positive curve) and off (negative curve). The dotted line in (a) represents where the magnitudes of these opposite displacements are equal. The inset shows a soliton with both the south-pole to north-pole vector and the direction of motion along the y-axis. (c,d) Experimental skyrmion velocity (c) with nonzero fill ratio, $U_1/U_2$, and (d) with $U_1/U_2 = 0$. The corresponding square waveform amplitude modulation schemes are shown in the inset. Chiral mixture is MLC-6609 and ZLI-811 where thickness and pitch are about 10 μm.

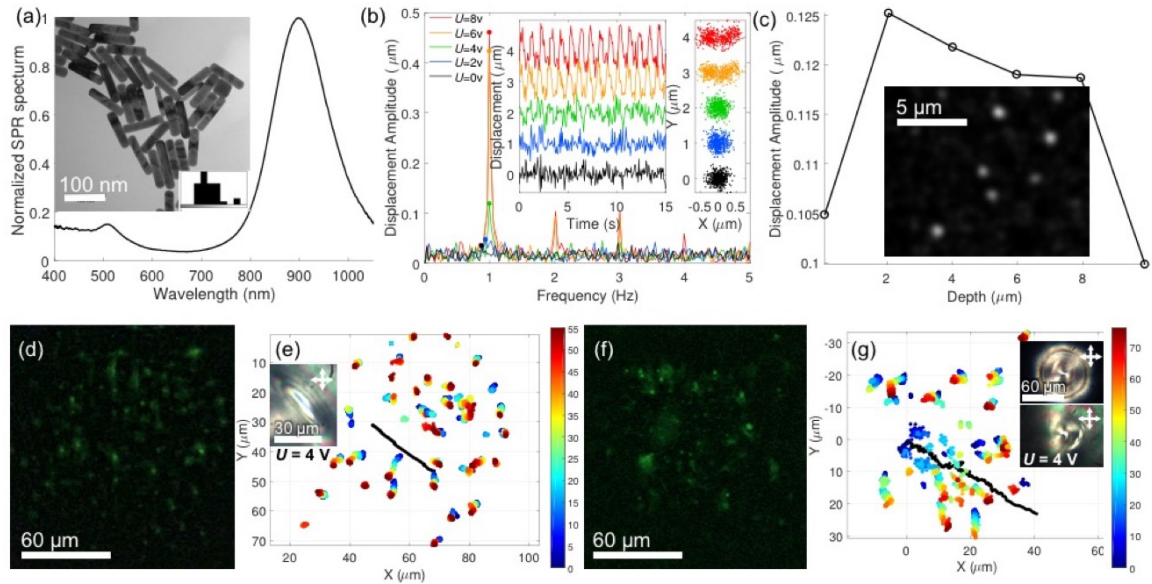

FIG. 12. Analysis of flows in LC cells at amplitude-modulated applied fields. (a) SPR spectrum of the GNR nanoparticles used as tracers of LC flow and TEM image (inset). (b) Fourier analysis of GNR motion for $U$ = 0 - 8 V, $f_c$ = 1 kHz, $f_m$ = 1 Hz and 50% duty cycle. Insets show displacement trajectories obtained for 15 cycles of amplitude modulation and the point cloud of nanoparticle positions for each trajectory offset in the y-direction for clarity. (c) Depth-resolved backflow displacement amplitude corresponding to Fourier component at 1 Hz in a 10 µm thick cell, $U$ = 5 V, $f_c$ = 1 kHz, $f_m$ = 1 Hz and 50% duty cycle and characteristic dark-field image of 7 GNRs at a depth of 4 µm (inset). (d) Fluorescence image of nanocubes in the midplane cross-section of a skyrmion's north-pole preimage in a 60-µm thick cell. (e) Time-coded trajectories of nanocubes during amplitude modulation for $U$ = 0 - 8 V, $f_c$ = 1 kHz, $f_m$ = 2 Hz and 50% duty cycle where the north pole traversed through nanocubes in the bulk LC (black, bottom right to top left). Inset POM image of the north-pole preimage and its vicinity. (f) Fluorescence images of nanocubes in the midplane cross-section of the north-pole and south-pole preimages of a skyrmion in a 60-µm thick cell. (g) Time-coded trajectories of nanocubes as the skyrmion traverses through (black line, top left to bottom right). The insets are POM images of the soliton before (top) and after (bottom) applying voltage. Chiral mixture is MLC-6609 and ZLI-811 where thickness and pitch are about 60 µm.

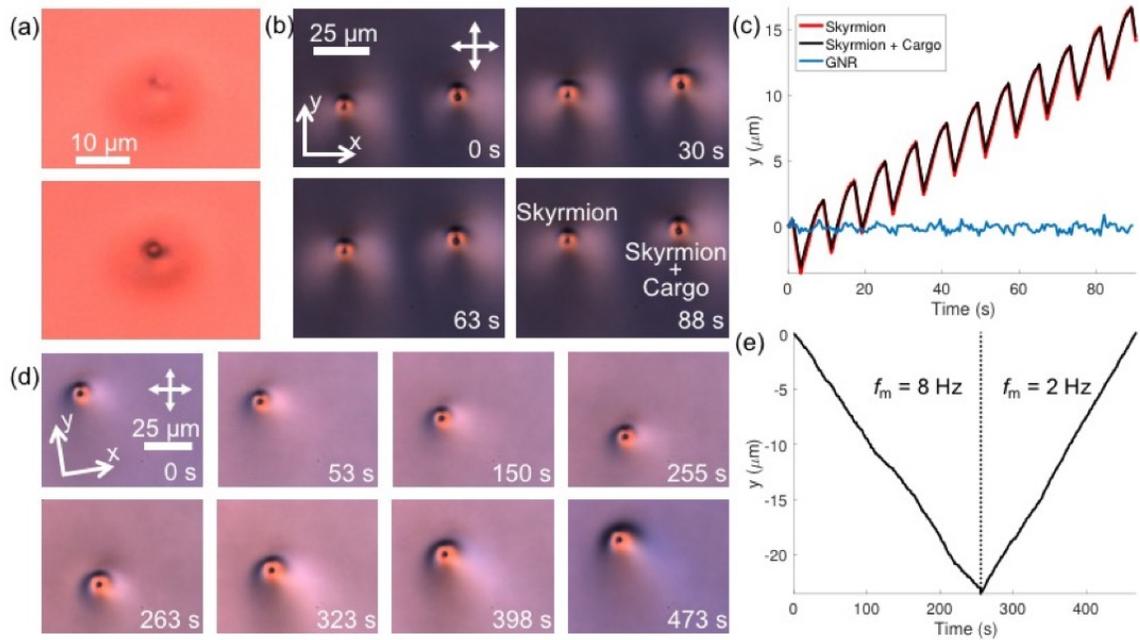

FIG. 13. Translational motion of soliton with and without entrapped melanin-resin microparticle cargo. (a) Bright-field microscopy of a skyrmion (top) and a skyrmion with cargo (bottom). (b) POM images of a skyrmion (left) and a skyrmion with cargo (right) during voltage modulation at $f_m$ = 2 Hz and 75% duty cycle. (c) Corresponding displacement of the skyrmion (red), skyrmion plus trapped cargo (black), and an untrapped GNR (blue). (d) POM images of a skyrmion transporting cargo during voltage modulation at $f_m$ = 8 Hz (top row) and $f_m$ = 2 Hz (bottom row). (e) Corresponding displacement of the skyrmion along y. Polarizer orientations are marked with white double arrows on the first frame and the elapsed time stamp is given at bottom-right of each POM frame. Chiral mixture is MLC-6609 and ZLI-811 where thickness and pitch are about 10 µm.

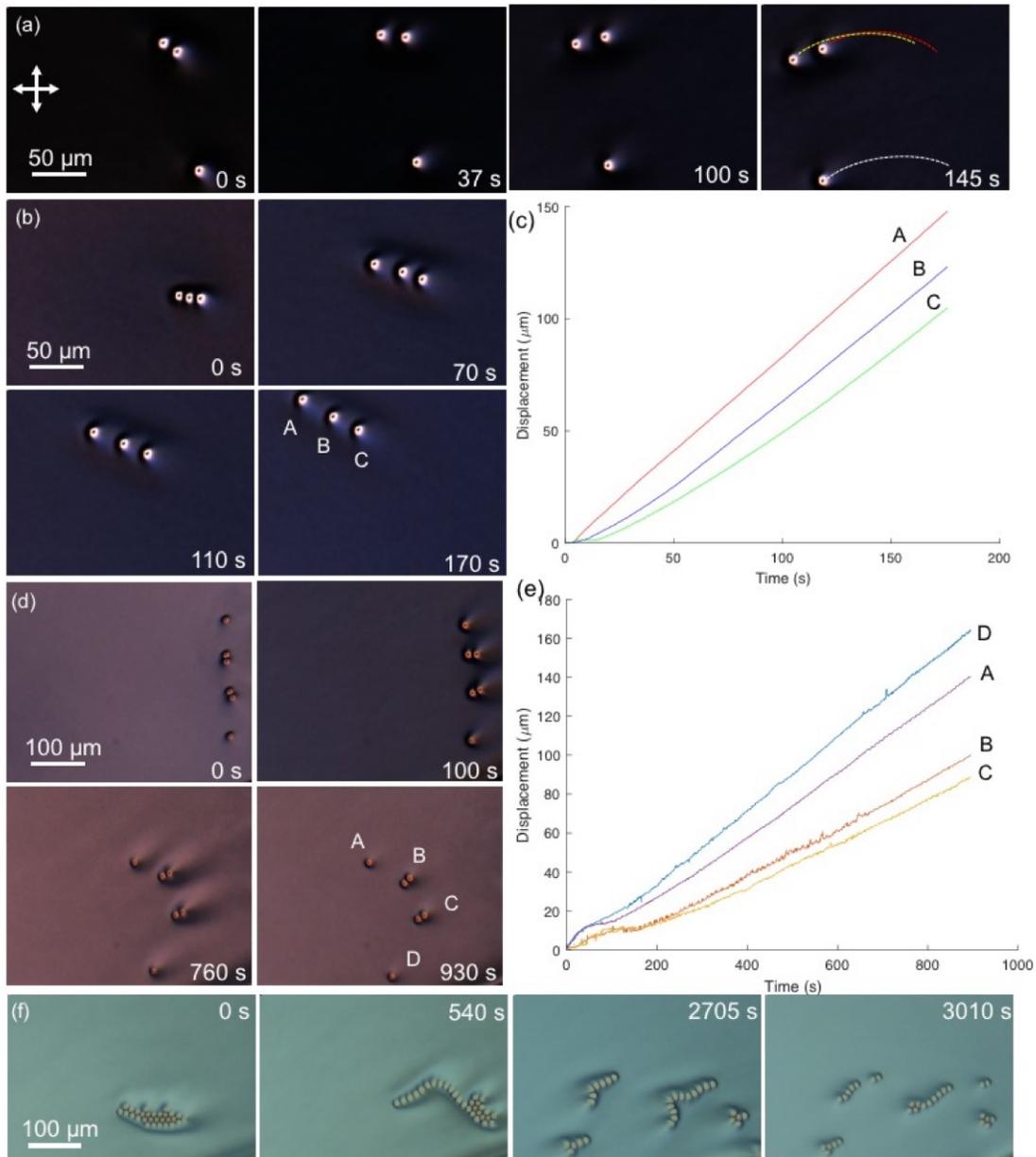

FIG. 14. Collective motion of skyrmions. (a) Single and paired motion in an arched path, with trajectories traced for each skyrmion on the last frame. (b) Motion of trio-chain and (c) corresponding displacement analysis where the three skyrmions are labeled A,B,C. (d) Single and paired motion race and (e) corresponding displacement analysis for two single skyrmions (labeled A,D) and two skyrmion pairs (labeled B,C). (f) Temporal evolution of the initially close-packed array of solitons and subsequent chain motion. White double arrows in (a) denote crossed polarizer orientation for all POM images and the elapsed time stamp is marked in the top- or bottom-right corner of each frame. Chiral mixture is MLC-6609 and ZLI-811 where thickness and pitch are about 10 μm.

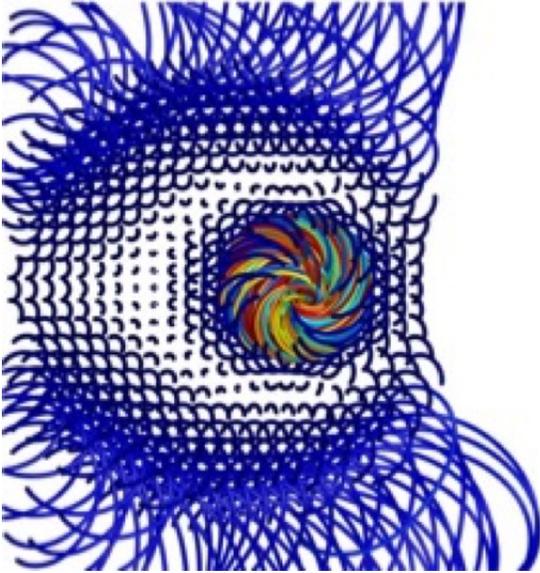

VIDEO 1. Video of director streamline evolution after voltage is removed in MLC-6609 with ZLI-811. The director structure relaxes from that of a toron embedded in TIC to the toron embedded in a uniform homeotropic background. The streamlines of the field originate from the midplane of the cell originally in equilibrium at $U = 4\text{V}$ and carrier frequency $f_\text{c} = 1$ kHz, followed by switching voltage to $U = 0$ V. The video was slowed down 8 times to reveal the details of temporal evolution of streamlines.

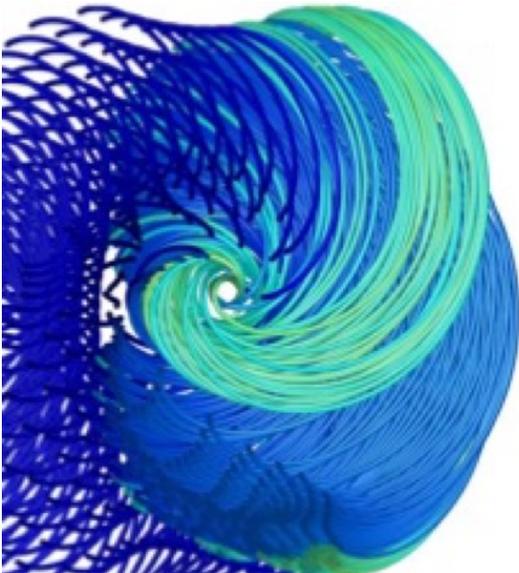

VIDEO 2. A video showing the non-reciprocal director streamline evolution in MLC-6609 with ZLI-811 during two periods of voltage driving with amplitude modulation ($U_1/U_2 = 0$, 75% duty cycle, $f_\text{m} = 2$ Hz, $f_\text{c} = 1$ kHz). The video was slowed down 2 times to reveal the details of temporal evolution of streamlines.

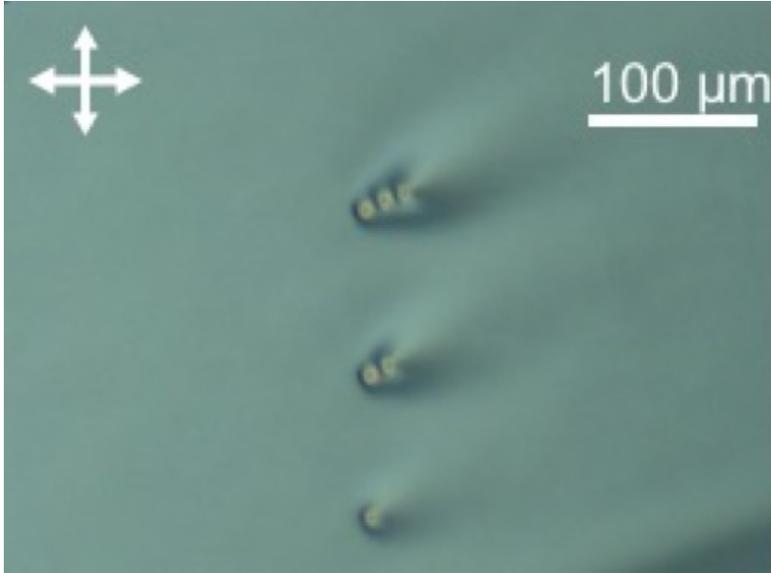

VIDEO 3. A POM video showing skyrmion chain motion race between a trio-chain, a pair, and a single skyrmion. The voltage-driving scheme parameters are: $U$ = 4.5V, 75% duty cycle, $f_c$ = 1 kHz, and $f_m$ = 2 or 8 Hz, as noted in top left of video frames. The video is sped up 10 times, and the total elapsed time is 29 minutes and 46 seconds. Chiral mixture is MLC-6609 and ZLI-811 where thickness and pitch are about 10 μm.

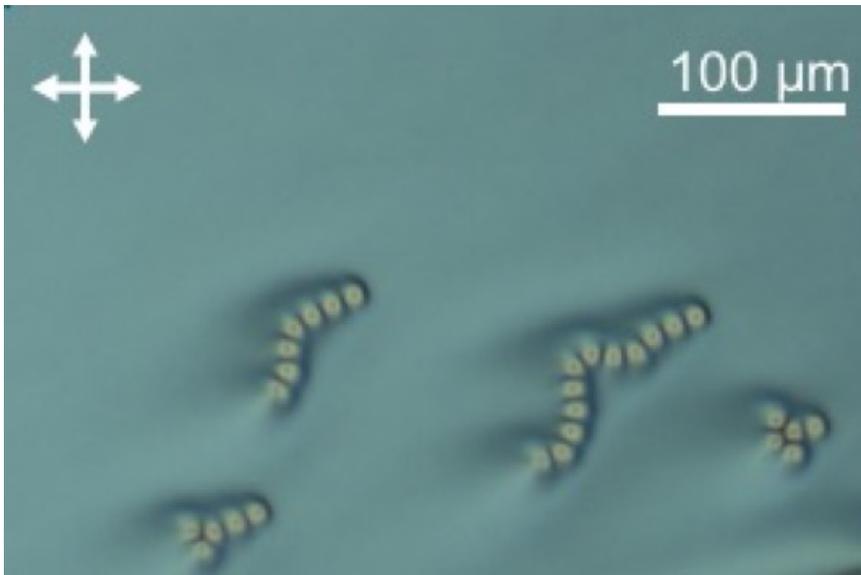

VIDEO 4. A video of kinetic out-of-equilibrium self-assembly and reassembly of skyrmion chains and other superstructures, which corresponds to frames in Fig. 14f. The voltage-driving scheme parameters are: $U$ = 4.5V, 75% duty cycle, $f_c$ = 1 kHz, and $f_m$ = 2 or 8 Hz, as noted in top left of video frames. The video is sped up 10 times, and the total elapsed time is 1 hour, 10 minutes, and 18 seconds. Chiral mixture is MLC-6609 and ZLI-811 where thickness and pitch are about 10 μm.

**Tables:**

**TABLE 1.** Material properties for used chiral nematic mixtures. The positive values of $h_{HTP}$ correspond to the right-handed chiral additives and the negative values correspond to the left-handed additive.

| Material/Property | E7 | MLC-6609 | ZLI-2806 |
|---|---|---|---|
| $\Delta\varepsilon$ | 13.8 | -3.7 | -4.8 |
| $h_{HTP}$ [additive] ($\mu m^{-1}$) | +7.3 [CB-15] | -10.5 [ZLI-811] | -8.3 [ZLI-811]<br>+5.9 [CB-15] |
| $K_{11}$ (pN) | 6.4 | 17.2 | 14.9 |
| $K_{22}$ (pN) | 3.0 | 7.5 | 7.9 |
| $K_{33}$ (pN) | 10.0 | 17.9 | 15.4 |